\documentclass[a4paper,12pt]{article}
\usepackage{jheppub,esint,shuffle,psfrag}
 \usepackage[utf8]{inputenc}
 
\usepackage{tikz}
\usetikzlibrary{arrows.meta,decorations.markings}

\usepackage{esint} 
\usepackage{breqn}
\usepackage{bm}
\usepackage{caption}
\usepackage{amssymb}
\usepackage{comment}

\usepackage{placeins}\input
\def \Tr {\mathop{\rm Tr}\nolimits}
\def \tr {\mathop{\rm tr}\nolimits}

\def \Re {\mathop{\rm Re}\nolimits}

\newcommand\lr[1]{{\left({#1}\right)}}

\newcommand \ket [1] {|{#1}\rangle}
\newcommand \bra [1] {\langle {#1}|}

\def \qqqquad {\qquad\qquad}

\def \ie {\textit{i.e.}\,}
\def \cf {\textit{cf.}\,}

\def\numberbysection{\@addtoreset{equation}{section}
                     \def\theequation{\thesection.\arabic{equation}}}

\begin{document}

\vspace*{0cm }

\author{Bercel Boldis$^{a,b}$, Dennis le Plat$^{b}$}
\affiliation{
$\null$ 
$^a$Department of Theoretical Physics, Institute of Physics, Budapest University of Technology and Economics M\H{u}egyetem rkp. 3., 1111 Budapest, Hungary
\\
$\null$
$^b$HUN-REN Wigner Research Centre for Physics, Konkoly-Thege Miklos ut 29-33, 1121 Budapest, Hungary

}
\title{From Fredholm Determinants to AdS/CFT Observables: A Universal Strong-Coupling Framework
}
\abstract{%
\small
We investigate the trans-series structure of the cusp anomalous dimension of $\mathcal{N}=4$~supersymmetric Yang--Mills theory and the dynamically generated mass gap of the $\mathrm{O}(6)$~sigma model at strong 't~Hooft coupling. We consider the one-parameter deformation of the cusp anomalous dimension, known as the tilted cusp, and review its strong-coupling expansion obtained through its representation in terms of Fredholm determinants with a matrix Bessel kernel. The advantage of this deformation is that the non-perturbative scales separate naturally, revealing structures that remain hidden in the physical limit. Motivated by this observation, we introduce the tilted mass gap, which reduces to the physical $\mathrm{O}(6)$~mass gap at a special value of the deformation parameter and exhibits a deep connection with the tilted cusp. This formulation allows us to determine the complete strong-coupling trans-series and resurgence structure of the $\mathrm{O}(6)$~mass gap, while extending and proving conjectured relations between successive non-perturbative corrections. Building on the underlying Fredholm determinant representation, we construct an Alien calculus that generates all non-perturbative sectors of these AdS/CFT observables directly from their perturbative expansions. Finally, we derive an exact all-orders relation between the strong-coupling trans-series of the $\mathrm{O}(6)$~mass gap and the cusp anomalous dimension of planar $\mathcal{N}=4$~supersymmetric Yang--Mills theory.
}

\maketitle

\section{Introduction and summary} \label{sec:Introduction}

The AdS/CFT~\cite{Maldacena:1997re,Gubser:1998bc,Witten:1998qj} correspondence provides a unique framework for studying both quantum field theory and string theory beyond the reach of conventional perturbative methods. In the planar limit~\cite{HOOFT1974461}, $\mathcal{N}=4$ supersymmetric Yang--Mills theory is integrable. Moreover, localization techniques~\cite{Pestun_2017} allow determining a wide class of observables exactly at any value of the 't Hooft coupling $\lambda$. This offers a rare opportunity to investigate the interplay between the weak- and strong-coupling regimes within the holographic duality.

Among the most important observables of $\mathcal{N}=4$ supersymmetric Yang--Mills theory is the \emph{cusp anomalous dimension}. More generally, the cusp anomalous dimension arises from the renormalization of Wilson loops containing a cusp, and its relation to the infrared asymptotics of gauge-theory amplitudes was established in early studies of QCD\footnote{We thank G. Korchemsky for bringing these references to our attention.}~\cite{Korchemsky:1985xj,Korchemsky:1987wg,Korchemsky:1991zp}. In the lightlike limit, the resulting quantity $\Gamma_{\mathrm{cusp}}(g)$ universally controls infrared singularities. It also governs the leading logarithmic growth of anomalous dimensions at large Lorentz spin, a relation first established from the $x \to 1$ behavior of QCD evolution kernels~\cite{Korchemsky:1988si}.

In planar $\mathcal{N}=4$ SYM, this large-spin behavior can be studied for single-trace operators of the form
\begin{equation}
    \mathcal{O}_{N,L}=\Tr\left\{D^{k_1}Z D^{k_2} Z\dots D ^{k_L} Z\right\}\,,
\end{equation}
with total spin $N=k_1+\dots +k_L$ and light-cone derivatives $D$ acting on complex scalar fields $Z$. In the limit in which N and L are large, but the ratio $j=L/\log N$ is kept fixed, the anomalous dimension of $\mathcal{O}_{N,L}$ at arbitrary 't Hooft coupling $g=\sqrt{\lambda}/(4\pi)$ scales as
\begin{equation} \label{eq:Gamma-AnomDim}
\gamma(g) = \left[2\Gamma_{\mathrm{cusp}}(g)+\epsilon(g,j)\right]\log N+\dots\,,
\end{equation}
where the function $\epsilon(g,j)$ is described by the ground state energy density $\epsilon_{\mathrm{O}(6)}$ of the $\mathrm{O}(6)$ sigma model
\begin{equation}
    \epsilon_{\mathrm{O}(6)}={\epsilon(g,j)+j\over 2}\,.
\end{equation}

In~\cite{Basso:2008tx} the cusp anomalous dimension was expressed in terms of the solution of the Beisert-Eden-Staudacher (BES) equation~\cite{Beisert:2006ez}, which describes the rapidity density of large-spin twist-two operators.
An advantage of this description is that it allowed the authors in~\cite{Basso:2009gh} to compute the leading perturbative and non-perturbative contributions for the cusp anomalous dimension at strong coupling by investigating the analytic properties of the solution of the BES equation. It was shown that the non-perturbative corrections scale with powers of
\begin{equation}\label{exp-scale-orig}
    \Lambda^2=\sigma(2\pi g)^{1/2}e^{-2\pi g}{\Gamma\left(\frac{3}{4}\right)\over \Gamma\left(\frac{5}{4}\right)}\,,\qquad \sigma=e^{-3i\pi/4}\,.
\end{equation}
The strong-coupling expansion and the resurgence of the first few non-perturbative sectors were further investigated in~\cite{Dorigoni:2015dha}.

The appearance of the $\mathrm{O}(6)$ mass gap in this context, and its explicit dependence on the 't Hooft coupling, originates naturally from the string theory dual of $\mathcal{N}=4$ SYM. In the strong-coupling regime (large $g$), the dynamics of these operators is governed by strings spinning in the $\mathrm{AdS}_5 \times \mathrm{S}^5$ background. In the Alday-Maldacena scaling limit~\cite{Alday:2007mf}, the low-energy effective field theory for the transverse string fluctuations on the $\mathrm{S}^5$ sphere reduces to an integrable two-dimensional $\mathrm{O}(6)$ non-linear sigma model. While this classical worldsheet theory is conformally invariant, it dynamically generates a mass gap $m_{\mathrm{O}(6)}$, which is exponentially small in the string tension. Because the string tension acts as the effective inverse coupling of the worldsheet theory and is directly proportional to the 't Hooft coupling $g$, the mass gap acquires a non-perturbative dependence on $g$. In fact, the square of this dynamical mass gap is directly proportional to the non-perturbative scale $\Lambda^2 \sim e^{-2\pi g}$ in~\eqref{exp-scale-orig}.

An important feature of the large-$g$ expansion of the cusp anomalous dimension is that the leading non-perturbative correction is related to the square of the $\mathrm{O}(6)$ mass gap~\cite{Basso:2008tx} through
\begin{equation}\label{cusp-mO6}
    \delta\Gamma_{\mathrm{cusp}}(g)=-{e^{-3i\pi/4}\over 4\sqrt{2}}m_{\mathrm{O}(6)}^2+O\left(m_{\mathrm{O}(6)}^4\right)\,.
\end{equation}
This remarkable statement directly probes the AdS/CFT correspondence, since it connects two quantities emerging from different regimes of the holographic duality. This relation was verified up to very high orders in $1/g$ in~\cite{Basso:2009gh,Dorigoni:2015dha} by expressing the mass gap in terms of the same set of functions as the cusp anomalous dimension, that is, the solution of the BES equation.

In~\cite{Dorigoni:2015dha} further important relations were derived between the different leading sectors of~$\Gamma_{\mathrm{cusp}}$ and~$m_{\mathrm{O}(6)}$. However, their full strong-coupling structure has so far been unknown, since relations such as (5.33)-(5.35) in~\cite{Dorigoni:2015dha} connecting the resurgence between the different non-perturbative sectors broke down at certain exponential orders. In~\cite{Boldis:2026rkb,Bajnok:2026xri} the authors showed that there is a universal governing structure for the non-perturbative regimes of the cusp anomalous dimension. The relation~\eqref{cusp-mO6} suggests that there is an even deeper relation between the strong-coupling expansion of~$\Gamma_{\mathrm{cusp}}$ and the~$\mathrm{O}(6)$ mass gap. It seems natural to wonder whether the universal structure revealed in~\cite{Bajnok:2025lji,Boldis:2026rkb} also holds for the mass gap~$m_{\mathrm{O}(6)}$. The aim of this paper is to demonstrate that this is indeed the case.

Beyond its universal role in infrared singularities and large-spin asymptotics, the cusp anomalous dimension also enters infrared-subtracted scattering amplitudes. In particular, in special kinematic limits of planar $\mathcal N=4$ SYM amplitudes, the finite part is also governed by a one-parameter deformation $\Gamma_{\mathrm{cusp}}(\alpha)$\footnote{From now on, for simplicity, we omit the dependence on the~'t~Hooft coupling.} of the cusp anomalous dimension, the so called \emph{tilted cusp anomalous dimension}, with $\Gamma_{\mathrm{cusp}}(\alpha=\pi/4)=\Gamma_{\mathrm{cusp}}$~\cite{Basso:2020xts,Basso:2022ruw}.

In a more precise manner, in~\cite{Basso:2020xts}, the tilted cusp was defined by the solution of a one-parameter deformation of the BES equation. It was written in terms of a two-by-two block matrix~$\mathbb{K}(\alpha)$ with each block being a semi-infinite matrix. The explicit form of~$\mathbb{K}(\alpha)$ is given by
\begin{equation}\label{K-alpha}
\mathbb K(\alpha)=2\cos\alpha \left[\begin{array}{rr}\cos\alpha\, K_{\text{oo}}  & \sin\alpha\,  K_{\text{oe}} \\[1.2mm] -\sin\alpha\,  K_{\text{eo}} & \cos\alpha\,  K_{\text{ee}} \end{array}\right]\,.
\end{equation}
The elements of the different blocks are parameterized as
\begin{equation}
\begin{aligned}\notag\label{blocks}
    [K_{oo}]_{n,m}&=K_{2n-1,2m-1}\,,\qquad &&[K_{oe}]_{n,m}=K_{2n-1,2m}\,,\\
    [K_{eo}]_{n,m}&=K_{2n,2m-1}\,,&&[K_{ee}]_{n,m}=K_{2n,2m}\,,
\end{aligned}
\end{equation}
with $K_{n,m}$ admitting the integral representation
\begin{equation}\label{K-ori}
K_{nm} ={1\over 2}\sqrt{nm}\int_0^\infty  {dt\over t} \chi\left(\frac{\sqrt{t}}{2g}\right)J_n(\sqrt{t}) J_m(\sqrt{t})\,,\qquad \chi(x)=\frac{2}{e^x-1}\,,
\end{equation}
where $J_n(x)$ is the Bessel function and $\chi(x)$ is called the \emph{symbol}. We see that the parameter $\alpha \in [0,\pi]$ is an angle controlling the contribution of the four semi-infinite blocks with even and odd parity. Then the tilted cusp is related to the resolvent of this matrix
\begin{equation}\label{tilted-cusp}
    \Gamma_{\mathrm{cusp}}(\alpha)=4g^2\left(1\over 1+\mathbb{K}(\alpha)\right)_{1,1} \,.
\end{equation}
This function evaluated at $\alpha=\pi/4$ coincides with the cusp anomalous dimension of $\mathcal{N}=4$ SYM.

In~\cite{Bajnok:2024bqr} an efficient method was presented to obtain the strong-coupling expansion for the cusp anomalous dimension. The method was based on the fact that, by using Cramer's rule, the tilted cusp can be written as the ratio of two determinants
\begin{equation}\label{tilted-cusp-det}
        \Gamma_{\mathrm{cusp}}(\alpha)=4g^2\frac{Z_{\ell=1}(\alpha)}{Z_{\ell=0}(\alpha)}\,,
    \end{equation}
with
\begin{equation}\label{F-def}
Z_\ell(\alpha) = \det\Big(\delta_{nm}+\mathbb K_{nm}(\alpha)\Big)\Big|_{1+\ell \le n,m<\infty}\,.
\end{equation}
Here $\ell\geq0$ is a non-negative integer. As shown in \cite{Bajnok:2024bqr}, \eqref{F-def} can be reformulated as Fredholm determinants, and the strong-coupling expansion of these determinants with arbitrary $a$ and $\ell$ can be efficiently computed through a set of integro-differential equations and analyticity conditions. Then, taking the ratio in~\eqref{tilted-cusp-det}, the large-$g$ expansion of $\Gamma_{\mathrm{cusp}}(\alpha)$ immediately follows.

An important property of the determinant \eqref{F-def} and hence the tilted cusp \eqref{tilted-cusp-det} with arbitrary mixing angle $\alpha$ is that if we move away from the physical value $\alpha=\pi/4$, two different exponential scales emerge in the strong-coupling expansion and consequently their large-$g$ trans-series become double sums involving powers of
\begin{equation} \label{eq:IntroduceNPScales}
    \Lambda_-^2\sim e^{-4 g(\pi-2\alpha)}\,,\qquad\Lambda_+^2\sim e^{-4 g(\pi+2\alpha)}\,.
\end{equation}
As it was first discussed in~\cite{Dunne:2025wbq}, the separation of these two exponential scales appear in the Borel analysis of the tilted cusp anomalous dimension and is necessary to build up a resurgence description for its strong-coupling trans-series. It is easy to see that for $\alpha=\pi/4$, from a certain order the exponential scales $\Lambda_-^2$ and $\Lambda_+^2$ merge, therefore several properties naturally appearing in the double expansion in $\Lambda_-^2$ and $\Lambda_+^2$, such as resurgence, remain hidden in the physical case.

Further investigating non-perturbative corrections of the \eqref{F-def}, the authors in~\cite{Boldis:2026rkb,Bajnok:2026xri} found a universal governing structure between different exponentially suppressed sectors for the determinant. It also made it possible to reveal the complete resurgence structure of the trans-series and develop a complete Alien algebra that connects the non-perturbative corrections. The same universal structure was previously found in~\cite{Bajnok:2025lji} for simpler determinants with a Bessel kernel, related to various $\mathcal N=2$ and $\mathcal{N}=4$ SYM observables. 

In this paper we continue the program of~\cite{Boldis:2026rkb,Bajnok:2026xri} and using the observations made for the determinant~\eqref{F-def} we further investigate the strong coupling structure and resurgence of the tilted cusp anomalous dimension. We also introduce a one parameter generalization of the~$\mathrm{O}(6)$ mass gap, referred to as the \emph{tilted mass gap}~$m(\alpha)$. It depends on the mixing angle~$\alpha$ and it yields the~$\mathrm{O}(6)$ mass gap when evaluated at~$\alpha=\pi/4$.  Its definition is conjectured, based on its relation with the cusp anomalous dimension at the physical value. We use the formalism developed for computing the tilted cusp anomalous dimension through the determinant~$Z_\ell(\alpha)$ and the related integro-differential equations to effectively compute the tilted mass gap. Introducing the mixing angle in~$m(\alpha)$ gives rise to the two naturally emerging exponential scales~\eqref{eq:IntroduceNPScales} (in contrast to only one in~\eqref{exp-scale-orig}) in its expansion at large values of the~'t~Hooft coupling. The appearance of two scales will allow us to explain why the relations found in~\cite{Dorigoni:2015dha} break down at a certain exponential order. Finally, following~\cite{Bajnok:2025lji,Boldis:2026rkb,Bajnok:2026xri}, the strong-coupling trans-series can be restructured and a universal resurgence structure and Alien calculus can be obtained for~$m(\alpha)$, and hence for the special case of the~$\mathrm{O}(6)$ mass gap.

The paper is structured as follows: In Section~\ref{Sec:tiltedCusp} we review the tilted cusp anomalous dimension. Starting from the BES equation~\cite{Beisert:2006ez,Eden:2006rx} and its solution, we discuss the cusp's formulation in terms of a ratio of Fredholm determinants afterwards. The powerful method of differential equations~\cite{Bajnok:2024epf,Bajnok:2024ymr,Bajnok:2024bqr} is reviewed in Section~\ref{Sec:MethodDiffEq}. We discuss two important sets of solutions to the integro-differential equations, namely far from the origin and at the origin. Combining these solutions in Section~\ref{SubSec:MatchingSols}, we can re-express the solution to the BES equation.

In Section~\ref{Sec:TiltedMassGap}, we define the tilted mass gap
as a one-parameter generalization of the physical
$\mathrm{O}(6)$ mass gap, motivated by the known relation between
the physical mass gap and the solution of the BES
equation~\cite{Alday:2007mf,Basso:2008tx,Bajnok:2008it,Basso:2009gh}.
We express this quantity in terms of the solutions to the generalized
BES equation and, equivalently, in terms of the solutions to the set
of integro-differential equations introduced in
Section~\ref{Sec:MethodDiffEq}.
We explicitly give its first few non-perturbative corrections
and discuss the physical $\mathrm{O}(6)$ case in
Section~\ref{Sec:O6-mass-gap-res}.

Finally, in Section~\ref{Sec:Universal-Structure} we turn to the universal strong-coupling structure of the observables considered. We start by discussing the determinant itself in Section~\ref{Sec:Universal-Resurgence-Determinant}, then the tilted cusp in Section~\ref{cusp-sec} and finally the tilted mass gap in Section~\ref{Sec:Tilted-mass-gap}. In particular, we work out their Alien algebra and derive general relations for ratios of their different non-perturbative sectors. The physical case of the$\mathrm{O}(6)$ mass gap is presented in Section~\ref{Sec:O6-mass-gap} for the reader's convenience.

We end with a discussion of our results in Section~\ref{Sec:Discussion}. The three Appendices~\ref{App:A},~\ref{Sec:AppB}, and~\ref{App:Proof} contain some further technical details.

\section{Tilted cusp} \label{Sec:tiltedCusp}

The \emph{tilted cusp anomalous dimension}~\cite{Basso:2020xts} was originally introduced as a one-parameter generalization of the $\mathcal{N}=4$ SYM cusp anomalous dimension. There are two different ways to obtain an expansion for the tilted cusp anomalous dimension at strong coupling: the first is by solving a one-parameter generalization of the BES equation in a way similar to that used in~\cite{Basso:2009gh}, or, second, as it was stated in~\eqref{tilted-cusp-det}, this quantity can be expressed as a ratio of two determinants, whose strong-coupling expansions can be computed through a set of integro-differential equations (see~\cite{Bajnok:2024bqr}). In this Section we briefly summarize the main ingredients of the two methods, first focusing on the one-parameter generalization of the BES equation and finally on the strong-coupling expansion of the determinants appearing in the ratio~\eqref{tilted-cusp-det}.

\subsection{Cusp from the BES equation} \label{Sec:Cusp-From-Bethe}

We begin with the computation of the tilted cusp from the BES equation.
We start by defining the auxiliary function
\begin{equation}\label{capital-G}
    \gamma^{(0)}(x)=i\frac{{\chi}_\alpha\left(x\right)}{\cos\left(\alpha\right)}\left(\gamma_+(x)+i\gamma_-(x)\right)\,,
\end{equation}
which is split into the parity even and odd functions $\gamma_\pm(x)$ with $\gamma_\pm(-x)=\pm \gamma_\pm(x)$. These functions are defined through the resolvent of the matrix $\mathbb{K}(\alpha)$ as
\begin{equation}
\begin{aligned}\label{gamma_pm_body}
    &\gamma_-(x)=\sum_{m\geq1}\sqrt{2m-1}J_{2m-1}(2gx)\left(\frac{1}{1+\mathbb{K}(\alpha)}\right)_{1,2m-1}\,, \\
    &\gamma_+(x)=\sum_{m\geq1}\sqrt{2m}J_{2m}(2gx)\left(\frac{1}{1+\mathbb{K}(\alpha)}\right)_{1,2m}\,.
    \end{aligned}
\end{equation}
Similar definitions were previously introduced in~\cite{Basso:2007wd}. The functions $\gamma_\pm(x)$ and hence $\gamma^{(0)}(x)$ also depend on the coupling constant $g$, but for simplicity we suppress this dependence in our notation. The tilted cusp anomalous dimension can be obtained from the function $\gamma^{(0)}(x)$ close to the origin by
\begin{equation} \label{tilted-cusp-gamma}
    \Gamma_{\mathrm{cusp}}(\alpha)=-2g\lim_{x\to 0}\gamma^{(0)}(x) \,.
\end{equation}
Under transformations $\alpha\to \alpha+\pi$ and $\alpha \to -\alpha$, $\Gamma_{\mathrm{cusp}}(\alpha)$ remains invariant. Thus, from now on, we restrict the values of the mixing angle to the interval $0\leq \alpha \leq \pi/2$.

The functions $\gamma_\pm(x)$ satisfy a system of integral equations involving the Bessel functions
\begin{equation}
\begin{aligned}\label{int-elem-zero}
    &\int{dy\over y}J_{2k-1}(2gy)\left[\left(1+\cos^2(\alpha)\,\chi\left(y\right)\right)\gamma_-(y)-\sin(\alpha)\cos(\alpha)\,\chi\left(y\right)\gamma_+(y)\right]={1 \over 2}\delta_{k,1} \,,\\
    &\int{dy\over y}J_{2k}(2gy)\left[\left(1+\cos^2(\alpha)\,\chi\left(y\right)\right)\gamma_+(y)+\sin(\alpha)\cos(\alpha)\,\chi\left(y\right)\gamma_-(y)\right]=0\,.
\end{aligned}
\end{equation}
These integral equations are one-parameter generalizations of the BES equation. It is straightforward to show that for $\alpha=\pi/4$ they reduce to equations (2.3) of~\cite{Basso:2009gh}. For a detailed derivation of these relations, starting from the definition~\eqref{gamma_pm_body}, we refer the reader to Appendix~\ref{App:A}.

The solution of~\eqref{int-elem-zero} can be derived in a similar way as for the $\alpha=\pi/4$ case. It can be shown that the auxiliary function $\gamma^{(0)}(x)$ can also be written in a form similar to the expression in~\cite{Basso:2009gh}, as
\begin{equation}\label{Gamma-sol-inteq}
    \gamma^{(0)}(x)=f_0(i2gx)V_0(i2gx)+f_1(i2gx)V_1(i2gx)\,,
\end{equation}
where the functions $f_0(x)$ and $f_1(x)$ are given by
\begin{equation}
\begin{aligned} \label{f01-sol}
    f_0(x)&=\sum_{n\geq 1}x\left[{d_+\over 4 \pi g n-x}U_1^+(4\pi g n)+{d_-\over 4\pi g n+x}U_1^-(4\pi g n)\right]-{1\over \sin2\alpha}\,, \\
    f_1(x)&=\sum_{n\geq 1}4\pi g n\left[{d_+\over 4\pi g n-x}U_0^+(4\pi g n)+{d_-\over 4\pi g n+x}U_0^-(4\pi g n)\right]\,.
\end{aligned}
\end{equation}
However, in our case, the coefficients $d_\pm$ also depend on the angle $\alpha$ and the coupling constant $g$, while the functions $U_k^\pm(x)$ and $V_k(x)$ are given by
\begin{equation}
\begin{aligned} \label{eq:IntVandU}
    V_0(x)&={2\sin\alpha\over \pi}\int_{-1}^1 dk e^{xk}{(1+k)^a\over (1-k)^a}=4ae^{-x}{}_1 F_1\left(a+1;2;2x\right)\,, \\
    V_1(x)&={2\sin\alpha\over \pi}\int_{-1}^1 dk e^{xk}{(1+k)^{a-1}\over (1-k)^a}=2e^{-x}{}_1 F_1\left(a;1;2x\right)\,,\\
    U_0^\pm(x)&={1\over 2}\int_{1}^\infty du e^{-x(u-1)}{(u-1)^{\pm a}\over (u+1)^{\pm a}}=\Gamma\left(1\pm a\right){e^x\over 2x}W_{\mp a,1/2}(2x)\,,\\
    U_1^\pm(x)&={1\over 2}\int_{1}^\infty du {e^{-x(u-1)}\over (u\mp 1)}{(u-1)^{\pm a}\over (u+1)^{\pm a}}={1\over 2}\Gamma\left({1\over 2}\pm \left(a-{1\over 2}\right)\right){e^x\over \sqrt{2x}}W_{\pm(1/2-a),0}(2x)\,.
\end{aligned}
\end{equation}
Here, ${}_1F_1(c;d;x)$ and $W_{c,d}(x)$ are the generalized hypergeometric and Whittaker functions, respectively. For simplicity, we have also introduced the notation $a=\alpha/\pi$, which we use interchangeably with the mixing angle $\alpha$ throughout the text. In the special case of $a=1/4$ these expressions agree with the ones presented in~\cite{Basso:2009gh}.

According to~\eqref{tilted-cusp-gamma} the tilted cusp anomalous dimension can then be obtained from the auxiliary function and is given by
\begin{equation}\label{gamma-fV}
    \Gamma_{\mathrm{cusp}}(\alpha)=-2g\left[f_0(0)V_0(0)+f_1(0)V_1(0)\right]\,.
\end{equation}

Following~\cite{Basso:2009gh,Dorigoni:2015dha}, the above solution can be used to compute the strong-coupling expansion of the cusp anomalous dimension. The method is based on the fact that the functions in~\eqref{gamma_pm_body} are analytic in $x$, and therefore~\eqref{capital-G} vanishes at the zeros of the generalized symbol $\chi_\alpha(x)$. The locations of these zeros are at $x=\pm 2\pi i x_j^\pm$ with
\begin{equation}\label{zeros}
     x_j^\pm=j+\frac12\mp a\,, \qquad \quad j\in \mathbb{N}_0\,.
\end{equation}
Substituting $x=\pm 2\pi i x_j^\pm$ into~\eqref{Gamma-sol-inteq} we arrive at an infinite set of quantization conditions
\begin{equation}\label{quant-fV}
    f_0(\mp4\pi g x_j^\pm)V_0(\mp4\pi g x_j^\pm)+f_1(\mp4\pi g x_j^\pm)V_1(\mp4\pi g x_j^\pm)=0\,,\qquad j\geq 0\,.
\end{equation}
This relation can be used to compute perturbative and non-perturbative coefficients for the functions $d_\pm$ in~\eqref{f01-sol} and to determine the strong-coupling expansion of the tilted cusp anomalous dimension from~\eqref{gamma-fV}. For the special value $a=1/4$, the quantization condition leads to the same leading order coefficients for $f_n(x)$ as in~\cite{Basso:2009gh}.

In the following, we present a more powerful technique to compute the large-$g$ expansion of the tilted cusp, which is based on the representation~\eqref{tilted-cusp-det}. However, it is important to summarize the method based on integral equations~\eqref{int-elem-zero} and their solutions, since the mass gap of the $\mathrm{O}(6)$ model can also be expressed in terms of the functions $f_n(x)$ and $V_n(x)$ with $a=1/4$. Later, this will play a crucial role in our derivation and discussion.

\subsection{Fredholm determinant}

The tilted cusp anomalous dimension can also be expressed as a ratio of two determinants, \cf~\eqref{tilted-cusp-det}. The strong-coupling coefficients of these determinants can be obtained in an efficient way through a set of integro-differential equations. The procedure can be found in detail in~\cite{Bajnok:2024bqr}. It was shown there that the cusp strong-coupling expansion can be written as a \emph{trans-series} of the form
\begin{equation} \label{Z-strong-Lambda}
    Z_\ell(\alpha)=A_\ell\sum_{n,m\geq0}\Lambda_-^{2n}\Lambda_+^{2m}Z^{(n,m)}\,.
\end{equation}
The non-perturbative scales, or \emph{trans-series parameters}, are defined as
\begin{equation}\label{Lambda-pm}
    \Lambda_-^2=e^{i\alpha}(8\pi g)^{2a}e^{-4\pi g(1-2a)}\,,\qquad\Lambda_+^2=e^{-i\alpha}(8\pi g)^{-2a}e^{-4\pi g(1+2a)}\,.
\end{equation}
The coefficient functions $Z^{(n,m)}$ are given by series in $1/g$. The perturbative part of the trans-series corresponds to the $n,m=0$ sector. The explicit value of the overall $g$-dependent prefactor $A_\ell$ can be found in (1.11) of~\cite{Boldis:2026rkb}. The series is normalized in such a way that the leading $1/g$ coefficient of the perturbative part is $1$. 

For practical reasons, let us state the first few coefficients of the perturbative sector of the determinant $Z_{\ell}(\alpha)$ as it plays an important role in the following. The perturbative expansion takes the form
\begin{equation} \label{eq:pertDet}
    Z_\ell^{(0,0)} = 1 + (a^2 - \ell^2) \sum_{k \ge 1} \frac{(-1)^k}{k!} \frac{f_k^{(0,0)}}{(8\pi g)^k}\,,
\end{equation}
with the coefficients given by
\begin{equation} \label{eq:pertDetCoeff}
    \begin{aligned}
        f_1^{(0,0)} &= \mathcal{I}_2\,, \\
        f_2^{(0,0)} &= (a^2 - \ell^2 + 1) \mathcal{I}_2^2 + 2a \mathcal{I}_3\,, \\
        f_3^{(0,0)} &= (a^2 - \ell^2 + 1)(a^2 - \ell^2 + 2) \mathcal{I}_2^3 + 6a (a^2 - \ell^2 + 2) \mathcal{I}_2 \mathcal{I}_3 + 2(5a^2 - \ell^2 + 1) \mathcal{I}_4 \,, \\
    f_4^{(0,0)} &= (a^2 - \ell^2 + 1)(a^2 - \ell^2 + 2)(a^2 - \ell^2 + 3) \mathcal{I}_2^4 + 12a (a^2 - \ell^2 + 2)(a^2 - \ell^2 + 3) \mathcal{I}_2^2 \mathcal{I}_3 +\\ &\phantom{==} 6(2a^4 - 2a^2 \ell^2 + 9a^2 - \ell^2 + 1) \mathcal{I}_3^2 + 8(a^2 - \ell^2 + 3)(5a^2 - \ell^2 + 1) \mathcal{I}_2 \mathcal{I}_4 + \\ &\phantom{==} 12a (7a^2 - 3\ell^2 + 5) \mathcal{I}_5.
    \end{aligned}
\end{equation}

The quantities $\mathcal{I}_n$, defined as
\begin{equation}\label{In-def}
\mathcal{I}_n={1\over\pi^2}\Re\left[ \int_0^\infty {dz} \,\lr{2\pi i\over z}^n z \partial_z \log \chi_\alpha(z) \right]\,,
\end{equation}
are called the \emph{moments} of the function and depend on the modified symbol
\begin{equation}\label{symb-gen}
\chi_\alpha(x)=e^{i\alpha}+\chi(x)\cos\alpha=\frac{\cosh(x/2+i\alpha)}{\sinh(x/2)}\,,
\end{equation}
with $\chi(x)$ given in~\eqref{K-ori}. Although the integrand in~\eqref{In-def} is divergent at the origin for $n\geq 2$, it can be evaluated using analytical regularization~\cite{Bajnok:2024qro}. The moment $\mathcal{I}_n$ can be expressed in terms of the polygamma functions $\psi^{(k)}(x)$ as 
\begin{equation}\label{In}
\mathcal{I}_{n}= {1\over (n-2)!}\bigg[ {} \psi ^{(n-2)}\left(\frac{1}{2}-a\right)-\psi ^{(n-2)}(1)  +(-1)^n \bigg(\psi ^{(n-2)}\left(\frac{1}{2}+a\right)- \psi ^{(n-2)}(1)\bigg) \bigg]\,.
\end{equation}
Finally, using the series representation of the polygamma function, the moments can be rewritten in terms of the poles and zeros of the symbol $\chi_\alpha(x)$, which yields
\begin{equation}
    \mathcal{I}_{n} = \sum_{k \geq 0} \left[ (-1)^{n-1} \left(\frac{1}{(x^+_k)^{n-1}} -\frac{1}{(y_k)^{n-1}} \right) - \left(\frac{1}{(x^-_k)^{n-1}} -\frac{1}{(y_k)^{n-1}} \right)  \right]\,,
\end{equation}
where $\chi_\alpha(x)$ has zeros at $x=\pm 2\pi i x_j^\pm$ and poles at $x=\pm 2\pi i y_j$, with $x_j^\pm$ given in \eqref{zeros} and $y_j=j+1$, $j\in \mathbb{N}_0$. Moreover, the moments satisfy the symmetry property
\begin{equation}\label{I-symm}
    \mathcal{I}_n(-a)=(-1)^n\mathcal{I}_n(a)\,.
\end{equation}

Finally, using~\eqref{eq:pertDet} and the explicit expression of the coefficient $A_\ell$  with $\ell=1$ and $\ell=0$, and expanding the ratio in~\eqref{tilted-cusp-det}, we obtain the perturbative part of the tilted cusp anomalous dimension as
\begin{equation}\label{tilt-cusp-pert}
    \Gamma_{\mathrm{cusp}}(\alpha)={8 g a\over\sin(2\alpha)}\left[1+\frac{\mathcal I_2}{8 \pi  g}-\frac{a\mathcal{I}_3}{(8\pi g)^2}+O\left(g^{-3}\right)\right]\,.
\end{equation}
Non-perturbative corrections obtained through \eqref{tilted-cusp-det} can be found by following the methods of~\cite{Bajnok:2024bqr,Boldis:2026rkb,Bajnok:2026xri}.

In~\cite{Boldis:2026rkb,Bajnok:2026xri} it was observed that for determinant observables the non-perturbative corrections can be recast into an infinite-parameter trans-series, in which each non-perturbative sector is related to the perturbative one. Similarly, iterative relations can be derived for the corresponding Stokes constants. These rules make the generation of non-perturbative corrections at strong coupling of the Fredholm determinants, and hence the tilted cusp anomalous dimension, particularly simple. We will give a more detailed review of the universal strong-coupling structure and resurgence relations for determinant observables in Section~\ref{Sec:Universal-Resurgence-Determinant}.

We can parameterize the strong-coupling expansion of the tilted cusp as
\begin{equation} \label{cusp-strong-Lambda}
    \Gamma_{\mathrm{cusp}}(\alpha)=\frac{8g a}{\sin (2\alpha)}\sum_{i,j\geq0}\Lambda_-^{2i}\Lambda_+^{2j}\Gamma^{(i,j)}\,,
\end{equation}
then the non-perturbative coefficient functions in the pure $\Lambda_-$-direction are related by
\begin{equation}\label{gamma-ratio-Lambdapm}
    {\Gamma^{(n+1,0)}\over \Gamma^{(n,0)}}={\Gamma^{(2,0)}\over \Gamma^{(1,0)}}\,,
\end{equation}
with $n\geq1$. We verified this remarkable observation at up to four non-perturbative orders in the $\Lambda_-$-direction and to the order $g^{-13}$. We will be able to prove the relation~\eqref{gamma-ratio-Lambdapm} in Section~\ref{cusp-sec} and present a generalization. 

Note that a similar relation was previously observed in~\cite{Dorigoni:2015dha} for the cusp anomalous dimension with $a=1/4$. However, this relation breaks down at the third non-perturbative order. This is due to the mixing of the two trans-series parameters~\eqref{Lambda-pm} for this specific value of $a$. It is customary to introduce the effective exponential scale as in~\cite{Basso:2009gh}
\begin{equation} \label{eq:EffectiveLambda}
    \Lambda^2=e^{-\frac{3i}{4}\pi }\frac{ \Gamma \left(\frac{3}{4}\right) }{\Gamma \left(\frac{5}{4}\right)}\sqrt{2 \pi  g} e^{-2 \pi  g}\,.
\end{equation}
The two trans-series parameters $\Lambda_+,\Lambda_-$ start to mix at $e^{-6\pi g}$ as
\begin{equation}\label{scale-mixture}
    \Lambda_-^2\sim\Lambda^2\,, \qquad \Lambda_-^4\sim\Lambda^4\,, \qquad \Lambda_-^6\,,\Lambda_+^2\sim\Lambda^6\,,\quad \dots
\end{equation}
Since the relation~\eqref{gamma-ratio-Lambdapm} only holds for the pure $\Lambda_-$-direction, it breaks down for the cusp anomalous dimension when $\Lambda_+$ starts to mix in.

\section{Method of differential equations} \label{Sec:MethodDiffEq}

Over the last few years, powerful techniques have been developed~\cite{Bajnok:2024epf,Bajnok:2024ymr,Bajnok:2024bqr} to efficiently calculate the strong-coupling expansion of Fredholm determinant observables, such as the cusp anomalous dimension, to high perturbative orders. Furthermore, this also allows access to non-perturbative corrections. 
These techniques were devised and tested in~\cite{Bajnok:2024bqr}. Recasting the matrix elements~\eqref{K-ori} into truncated Bessel operators, a set of integro-differential equations related to the observable can be found~\cite{Belitsky:2019fan,Belitsky:2020qrm,Belitsky:2020qir,Bajnok:2024bqr}. In the following, we recall the ingredients and the recipe for this method, as they will play an important role in the calculation of the $\mathrm{O}(6)$ mass gap. More details on the derivation are given in~\cite{Bajnok:2024bqr} and Appendix~\ref{App:A} of this paper. 

Our starting point is the function $\gamma^{(0)}(x)$ defined in~\eqref{capital-G}. In Section~\ref{Sec:tiltedCusp} we discussed how the tilted cusp anomalous dimension is related to the behavior of the auxiliary functions $\gamma^{(0)}(x)$ close to the origin, \cf~\eqref{tilted-cusp-gamma}. In fact, these functions are related to another set of functions $q_\pm$ through
\begin{equation}
\begin{aligned}\label{gamma-q-rel2}
    \gamma^{(0)}(x)=\frac{e^{i\alpha}}{4 x}\frac{\chi_\alpha(x)}{\cos\alpha}&\left[q_+(0)\partial_gq_-(x)-\partial_gq_+(0)q_-(x)+\right.\\
    &\left.+q_-(0)\partial_gq_+(x)-\partial_gq_-(0)q_+(x)\right]\,.
\end{aligned}
\end{equation}
For the proof of the relation and the precise definition of the functions $q_\pm(x)$, we refer the reader to Appendix~\ref{App:A} of this paper and to Appendix A and B of~\cite{Bajnok:2024bqr}. In~\cite{Bajnok:2024bqr} these functions were used to develop an efficient method for the strong-coupling expansion of the determinants~\eqref{F-def}. The aim is to express $m_{\mathrm{O}(6)}$ in terms of these well-known functions. 

In the following we summarize the relations that are sufficient to compute the strong-coupling expansion of the functions $q_\pm(x)$. However, it is important to note that we need to distinguish the solutions for the set of integro-differential equations far from and close to the origin. Although the equations are well-defined everywhere, the solution we find in the bulk diverges at the origin. In Section~\ref{sec:Far-from-Orig} we will therefore discuss the solution far from the origin, which was also the case studied in~\cite{Bajnok:2024bqr}. For the tilted cusp, however, we can see from~\eqref{tilted-cusp-gamma} that the behavior close to the origin becomes important. Hence, in Section~\ref{sec:Close-to-Orig} we study the corresponding solution to the set of integro-differential equations at the special point $x=0$.

\subsection{Functions away from the origin} \label{sec:Far-from-Orig}

We introduce the auxiliary functions $q_\pm(x)$ as well as the functions $W_0$, $W_\pm$. These functions satisfy a set of integro-differential equations that were derived in~\cite{Bajnok:2024bqr}. The differential equations are given by
\begin{equation}\label{diffeq-q-arbx}
    \left[(g\partial_g)^2 +(2gx)^2+W_0(g)\right]q_\pm(x)=W_\mp(g)q_\mp(x)\,,
\end{equation}
and the integro-differential equations read as follows
\begin{equation}
    \begin{aligned}\label{intrel-Wq}
& \partial_g W_0=-4 g \partial_g\left[g \cos \alpha \int_0^{\infty} d x \operatorname{Re}\left(q_{+}(x) q_{-}(x)\right) x^2 \partial_x \chi(x)\right]\,, \\
& \partial_g W_{-}=4 g \partial_g\left[g \cos \alpha \int_0^{\infty} d x \operatorname{Re}\left(q_{+}^2(x)\right) x^2 \partial_x \chi(x)\right]\,, \\
& \partial_g W_{+}=4 g \partial_g\left[g \cos \alpha \int_0^{\infty} d x \operatorname{Re}\left(q_{-}^2(x)\right) x^2 \partial_x \chi(x)\right] \,.
\end{aligned}
\end{equation}

At strong coupling, the ansatz for the auxiliary functions $q_\pm$ is given by~\cite{Bajnok:2024epf}
\begin{equation}
\begin{aligned}\label{q-ansatz}
& q_{+}(x)= \frac{g^{-a}}{\sqrt{2 \pi g \cos \alpha}}\left[\frac{(g x)^a e^{2 i g x}}{\Phi_+(x)} a_{+}(i x)+ \frac{(g x)^{-a} e^{-2 i g x}}{\Phi_{-}(-x)} b_{-}(i x)\right]\,, \\
& q_{-}(x)=\frac{g^a}{\sqrt{2 \pi g \cos \alpha}}\left[\frac{(g x)^{-a} e^{-2 i g x}}{\Phi_{-}(-x)} b_{+}(i x)+ \frac{(g x)^a e^{2 i g x}}{\Phi_+(x)} a_{-}(i x)\right]\,.
\end{aligned}
\end{equation}
Here, we introduce the functions $\Phi_\pm(x)$ that are related to the Wiener-Hopf-type decomposition of the generalized symbol function~\eqref{symb-gen} as
\begin{equation}
\chi_\alpha(x)=\frac{2\cos\alpha}{x}\Phi_+(x)\Phi_{-}(-x)\,.
\end{equation}
The functions $\Phi_+(x)$ and $\Phi_-(-x)$ are analytic in the lower and upper half-plane, respectively. Moreover, $\Phi_\pm(x)$ can be written in terms of their respective sets of zeros, characterizing the analytic structure. Explicitly, they take the following form~\cite{Bajnok:2024epf}
\begin{equation}
\begin{aligned}
{}& \Phi_+(x)=\frac{\Gamma \left(\frac12-a\right) \Gamma \left(1+\frac{i x}{2 \pi }\right)}{\Gamma \left(\frac12-a+{ix\over 2\pi}\right)}=\prod_{j\geq0}\frac{1+\frac{ix}{2\pi x_j^+}}{1+\frac{ix}{2\pi y_j}}\,,\\
{}& \Phi_-(x)=\frac{\Gamma \left(\frac12+a\right) \Gamma \left(1+\frac{i x}{2 \pi }\right)}{\Gamma \left(\frac12+a+{ix\over 2\pi}\right)}=\prod_{j\geq0}\frac{1+\frac{ix}{2\pi x_j^-}}{1+\frac{ix}{2\pi y_j}}\,,
\end{aligned}
\end{equation}
with $x_j^\pm$ and $y_j$ are given in \eqref{zeros}.

In addition, the coefficient functions $a_\pm(ix)$, $b_\pm(ix)$ were introduced in~\eqref{q-ansatz}. At strong coupling, these are given by trans-series
\begin{equation}\label{apm-bpm}
    a_\pm(ix) =\sum_{n,m \geq 0}\Lambda_-^{2n} \Lambda_+^{2m}a^{(n,m)}_{\pm}(ix)\,, 
    \qquad
    b_\pm(ix) =\sum_{n,m \geq 0}\Lambda_-^{2n} \Lambda_+^{2m}b^{(n,m)}_{\pm}(ix)\,. 
\end{equation}
The coefficient functions $a^{(n,m)}_{\pm}(ix)$ and $b^{(n,m)}_{\pm}(ix)$ denote the $(n,m)$-th non-perturbative sector. Each coefficient function itself has a perturbative expansion in $1/g$.

In Appendix~\ref{App:A} we discuss how the functions $q_\pm(x)$ satisfy certain symmetry properties under the exchange $a\to -a$, \cf~\eqref{q-pma-conj}. As a consequence, $a_\pm(ix)$ and $b_\pm(ix)$ satisfy
\begin{equation}\label{a-b-pma}
    b^{(n,m)}_\pm(ix|\alpha)=a^{(m,n)}_\pm(-ix|-\alpha)\,.
\end{equation}

Similarly, the trans-series ansatz for the functions $W_0$, $W_\pm$ is
\begin{equation}
\begin{aligned}\label{W-trans}
{}& W_0 = W_0^{(0)} +g^2\sum_{n+m > 0}\Lambda_-^{2n} \Lambda_+^{2m} \, W_0^{(n,m)}\,, \qquad 
\\
{}& W_\pm  = g^{\pm 2a} \bigg[ W_{\pm}^{(0)} + g^2\sum_{n+m > 0}\Lambda_-^{2n} \Lambda_+^{2m} \, W_{\pm}^{(n,m)}\bigg]\,,
\end{aligned}
\end{equation}
where the perturbative and non-perturbative coefficient functions were separated to have their respective large-$g$ expansion beginning at $O(g^0)$. As worked out in Appendix~\ref{App:A}, the functions $W_0^{(n,m)}$ and $W_\pm^{(n,m)}$ satisfy the symmetry properties
\begin{equation}\label{W-symm}
    W_0^{(n,m)}(\alpha)=W_0^{(m,n)}(-\alpha)\,,\qquad W_-^{(n,m)}(\alpha)=W_+^{(m,n)}(-\alpha)\,,
\end{equation}
where $W_0^{(0,0)}$ is understood as the perturbative part $W_0^{(0)}$.

Inserting the ansätze \eqref{q-ansatz}, \eqref{apm-bpm}, and \eqref{W-trans} into the system of integro-differential equations, the coefficients can be obtained by solving the set of relations order by order. In this way, one finds the leading perturbative coefficients for the functions $a_\pm(ix)$ as
\begin{equation}
\begin{aligned}\label{apm-sol}
    a^{(0,0)}_+(ix)&=1+\frac{ a^2}{4 g ix}-\frac{a^2 \left(2\pi(a-1)^2-2 i x \mathcal I_2\right)}{64\pi g^2 x^2}+O\left(g^{-3}\right)\,,\\
    a^{(0,0)}_-(ix)&={4^{2 a-1}\over gix} \frac{\Gamma (1-a)}{\Gamma (a)}\left[1+\frac{2\pi(a-1)^2+ (2 a-1) ix \mathcal I_2}{8\pi gi x }+O\left(g^{-2}\right)\right]\,.
\end{aligned}
\end{equation}
For $W_0$ and $W_+$ we have
\begin{equation}
\begin{aligned}
    W^{(0)}_0&=\left(a^2-\frac{1}{4}\right)-{ 2 a^2 \mathcal I_2\over 8\pi g}+{3 a^2 \left(2 a \mathcal I_3+\mathcal I_2^2\right)\over (8\pi g)^2}+O\left(g^{-3}\right)\,,\\
    W^{(0)}_+&= \frac{4^{2 a} (2a-1)\Gamma (1-a)}{\Gamma (a)}\times\\
    &\times\left[1+{2(a-1) \mathcal I_2\over 8\pi g} -\frac{(2 a-3) \left((a-1) \left((2 a-1) \mathcal I_2^2-3 a\mathcal I_3\right)-\mathcal I_3\right)}{(1-2a)(8\pi g)^2}+O\left(g^{-3}\right)\right]\,.
\end{aligned}
\end{equation}
Further coefficients can be obtained using~\eqref{a-b-pma} and~\eqref{W-symm} or can be found in the literature~\cite{Bajnok:2024bqr}. 
Since fixing the leading-order coefficients requires more sophisticated methods, for simplicity we only present their analytic expressions above, and for a more detailed discussion, we refer the reader to~\cite{Bajnok:2024bqr}.

Notice that in the $x\to 0$ limit, the coefficient functions in~\eqref{apm-sol} diverge; therefore the strong-coupling ansatz in~\eqref{q-ansatz} is only valid away from the origin. As mentioned above, the behavior at the origin plays an important role in the construction of the strong-coupling expansion of the function~$\gamma_\pm(x)$ and hence for the cusp anomalous dimension and the$\mathrm{O}(6)$ mass gap.

Equipped with the relations and initial values above, the perturbative part of the functions $q_\pm(x)$, $W_0$ and $W_\pm$ can be efficiently worked out up to arbitrarily high orders. Setting the trans-series parameters $\Lambda_\pm \to 0$ and solving~\eqref{diffeq-q-arbx} leads to an expression for the perturbative coefficients $a_{\pm}^{(0,0)}$, $b_{\pm}^{(0,0)}$ in terms of the coefficients of $W_0^{(0)}$, $W_\pm^{(0)}$. The integro-differential equations~\eqref{intrel-Wq} then allow us to express the coefficients of $W^{(0,0)}$ in terms of the moments $\mathcal{I}_n$ from~\eqref{In-def}.

In principle a similar procedure could be followed for the non-perturbative correction at a given order in the trans-series parameters $\Lambda_\pm$. However, solving the differential equation~\eqref{diffeq-q-arbx} at non-perturbative orders, one finds that the functions $a^{(n,m)}_\pm(ix)$ and $b^{(n,m)}_\pm(ix)$ produce poles at certain values of $x=x_\star$, which are related to the zeros of the symbol function $\chi_\alpha(x)$. This contradicts the fact that $q_\pm(x)$ is an analytic function of $x$. Therefore, instead of solving~\eqref{intrel-Wq} directly, we can employ the \emph{quantization condition} (similar to~\eqref{quant-fV}), which requires that $q_\pm(x)$ are free of spurious poles at finite $x=x_\star$, namely
\begin{equation} \label{qc0}
\lim_{x\to x_\star} (x-x_\star) q_\pm(x) = 0 \,.
\end{equation}
This allows us to relate the non-perturbative coefficients to the ones at lower orders and generate exponentially suppressed corrections in an efficient way. For a detailed discussion, see~\cite{Bajnok:2024qro,Bajnok:2024bqr}.

Finally, the solution $W_0$ can be used to find the observable $Z_\ell(\alpha)$ through~\eqref{eq:Z-and-W}.

\subsection{Functions at the origin} \label{sec:Close-to-Orig}

In the previous subsection, we reviewed how to compute the functions $q_\pm(x)$ for finite $x$. However, attempting to naively take the limit $\lim_{x \to 0} q_{\pm}(x, g)$ fails; the ansätze in \eqref{apm-bpm} for $a_\pm$, $b_\pm$ diverge at the origin (see, e.g., the leading coefficients in \eqref{apm-sol}). Therefore, a slightly different approach is necessary to determine $q_\pm(x)$ at $x=0$.

Naturally, we can solve~\eqref{diffeq-q-arbx} directly at the origin $x=0$ to obtain $q_\pm(0)$. The differential equation becomes
\begin{equation} \label{eq:DiffEq-c}
    (g\partial_g)^2q_\pm(0)=-W_0 q_\pm(0)+W_\mp q_\mp(0)\,.
\end{equation}

From the definition of the functions $q_\pm(x)$ it follows that at the origin they are finite; hence the combinations inside the brackets on the right-hand side of \eqref{q-ansatz} should also be finite at $x=0$. In this spirit, we can search for $q_\pm(0)$ of the form
\begin{equation}\label{q0-ansatz}
    q_\pm(0)={g^{\mp a}\over \sqrt{2\pi g\cos\alpha}}e^{-i\alpha/2}c_\pm\,,
\end{equation}
where $c_\pm(g)$ is again a trans-series in the coupling $g$ and the two exponential scales from \eqref{Lambda-pm}
\begin{equation}\label{cpm}
    c_\pm=\sum_{n,m \geq 0}\Lambda_-^{2n} \Lambda_+^{2m}c^{(n,m)}_{\pm}\,.
\end{equation}
The coefficient functions $c^{(n,m)}_{\pm}$ have a series expansion in $1/g$. 

The only ingredients missing to compute $q_\pm(0)$ at strong coupling is one of the initial values of the $O(1)$ coefficients either in $c_+^{(0,0)}$ or $c_-^{(0,0)}$. Once this value is fixed, the differential equation~\eqref{eq:DiffEq-c} can be solved order by order to obtain any $1/g$ contribution up to arbitrary exponential order both in $c_+$ and $c_-$. This initial value can be determined by differentiating~\eqref{gamma-q-rel2} with respect to $g$. Using the differential equation in~\eqref{diffeq-q-arbx}, we find that
\begin{equation}
    \partial_g \left[g\,\gamma^{(0)}(x)\right]=-gxe^{i\alpha}\frac{\chi_\alpha(x)}{\cos(\alpha)}\left[q_+(0)q_-(x)+q_-(0)q_+(x)\right]\,.
\end{equation}
Taking the limit $x\to 0$ and using~\eqref{tilted-cusp-gamma} we obtain a relation between $q_\pm(x)$ at the origin and the tilted cusp anomalous dimension
\begin{equation}\label{qpm-cusp}
    q_+(0)q_-(0)={e^{-i\alpha}\over 8g}\partial_g\Gamma_{\mathrm{cusp}}(\alpha)\,,
\end{equation}
which allows us to determine the initial value up to an overall sign. Putting together~\eqref{eq:DiffEq-c} and~\eqref{qpm-cusp}, the perturbative part of $c_+$ reads as
\begin{equation} \label{cp-sol}
    c^{(0,0)}_+(g)=\pm 4^{-a}\Gamma(1+a)\left[1-\frac{a \mathcal{I}_2}{8\pi g}+\frac{a \left((a+1) \mathcal{I}_2^2+(3 a+1) \mathcal{I}_3\right)}{2 (8\pi g)^2}+O\left(g^{-3}\right)\right]\,.
\end{equation}
 The perturbative series $c_-^{(0,0)}$ can be obtained from this expansion through the symmetry property
\begin{equation}
    c_-(\alpha)=c_+(-\alpha).
\end{equation}
Notice that the overall sign ambiguity is not fixed by condition~\eqref{qpm-cusp}. Although fixing this ambiguity lacks a rigorous proof, later we will see that only the upper sign gives the correct, physical result for the mass gap. Hence, we will only use the upper sign hereafter in order not to clutter the notation.

The method of differential equations~\eqref{eq:DiffEq-c} can be applied to determine the non-perturbative series~$c_\pm^{(n,m)}$. It turns out that after finding the leading $1/g$ coefficient in $c_\pm^{(0,0)}$, the whole strong-coupling expansion of $c_\pm$ is fixed. In Appendix~\ref{Sec:AppB} we present some explicit expressions for non-perturbative coefficients.

\subsection{Matching the solutions} \label{SubSec:MatchingSols}

Finally, we combine the above results to give a strong-coupling ansatz for the function $\gamma^{(0)}(x)$. Then, using the solution~\eqref{Gamma-sol-inteq} of the generalized BES equations~\eqref{int-elem-zero}, we can express the functions $f_n(x)$ and $V_n(x)$ in terms of the well-known strong-coupling functions $a_\pm(ix)$ and $c_\pm$.

Starting from the expression~\eqref{gamma-q-rel2} for $\gamma^{(0)}(x)$ and substituting the ansätze~\eqref{q-ansatz} and~\eqref{q0-ansatz} we can bring it to the form
\begin{equation}\label{Gamma-sol-AB}
\begin{aligned}
    \gamma^{(0)}(x)=e^{2igx}{(igx)^{a-2}\over 4\pi \cos\alpha}\Phi_-(-x)&\left[\left(g\partial_gc_+\right)a_-(ix)-c_+\left((3a+2igx)a_-(ix)+g\partial_ga_-(ix)\right)+\right.\\
    +&\left(g\partial_gc_-\right)a_+(ix)-c_-\left((-a+2igx)a_+(ix)+g\partial_ga_+(ix)\right)\left.\right]+\\
    +&\left(x\to-x,a\to-a\right)\,.
\end{aligned}
\end{equation}
Note that the last term becomes proportional to $e^{-2igx}$ after exchanging $x\to -x$ and $a\to -a$. Remarkably, this allows us to express $\gamma^{(0)}(x)$ purely in terms of the previously discussed and known non-perturbative quantities $a_\pm(ix)$, $b_\pm(ix)$ and $c_\pm$.

In the next step, let us compare the above expression~\eqref{Gamma-sol-AB} to the solution of the integral equations~\eqref{Gamma-sol-inteq}. 
The following useful identity (valid for $\mathrm{Re}\,x\geq0$, $x\neq 0$) allows us to relate the integrals from~\eqref{eq:IntVandU} with the functions $a_\pm(ix)$ and $c_\pm$ as
\begin{equation}
\begin{aligned}\label{eq:U-V-rel}
    V_0(ix)&=-{4\sin\alpha\over \pi}e^{i\alpha}\left(e^{ix}U_0^-(-ix)+e^{-ix}U_0^+(ix)\right)\,, \\
    V_1(ix)&=-{4\sin\alpha\over \pi}e^{i\alpha}\left(e^{ix}U_1^-(-ix)-e^{-ix}U_1^+(ix)\right)\,.
\end{aligned}
\end{equation}
Substituting these expressions into~\eqref{Gamma-sol-inteq}, the solution becomes
\begin{equation}
\begin{aligned}\label{eq:Gamma-f-and-U}
    \gamma^{(0)}(x)=&-{4\sin\alpha\over \pi}e^{i\alpha}e^{2igx}\left[f_0(i2gx)U^-_0(-i2gx)+f_1(i2gx)U_1^-(-i2gx)\right]-\\
    &-{4\sin\alpha\over \pi}e^{i\alpha}e^{-2igx}\left[f_0(i2gx)U^+_0(i2gx)-f_1(i2gx)U_1^+(i2gx)\right]\,.
\end{aligned}
\end{equation}
By comparing the functions multiplying the oscillating factors $e^{\pm 2igx}$ with those appearing in~\eqref{Gamma-sol-AB}, we can directly relate the strong-coupling ansätze with the solution of the generalized BES equations, which yields
\begin{equation}
\begin{aligned}\label{fU-ac}
    f_0(i2gx)U^-_0(-i2gx)+&f_1(i2gx)U_1^-(-i2gx)=\\
    ={(-igx)^{a}\over 2(2gx)^2 \sin\left(2\alpha\right)}\Phi_-(-x)&\left[\left(g\partial_gc_+\right)a_-(ix)-c_+\left((3a+2igx)a_-(ix)+g\partial_ga_-(ix)\right)+\right.\\
    +&\left(g\partial_gc_-\right)a_+(ix)-c_-\left((-a+2igx)a_+(ix)+g\partial_ga_+(ix)\right)\left.\right]\,.
\end{aligned}
\end{equation}
valid for $\mathrm{Re}\,x\geq0$, $x\neq 0$. This identity will play a crucial role in the following section. At perturbative order, we can verify this relation by inserting the leading order solutions for $a_\pm(ix)$, for $c_\pm$ from~\eqref{cp-sol} and for $f_n(i2gx)$ obtained from the quantization condition~\eqref{quant-fV}. Recall that the solution~\eqref{cp-sol} had a sign ambiguity. However, here we find that at strong coupling only the upper sign matches~\eqref{eq:Gamma-f-and-U}.

\section{Tilted mass gap} \label{Sec:TiltedMassGap}

Let us now turn to the mass gap of the $\mathrm{O}(6)$ model. In~\cite{Basso:2008tx,Basso:2009gh} the authors expressed the mass gap~$m_{\mathrm{O}(6)}$ in terms of the same functions $f_n(x)$ and $V_n(x)$ that appeared for the cusp anomalous dimension.
Moreover, there exists a one-parameter deformation of the cusp anomalous dimension, which can also be expressed in terms of some special functions. These functions reduce to the same $f_n(x)$ and $V_n(x)$ of~\cite{Basso:2008tx,Basso:2009gh} for the special value of the mixing angle $a=1/4$. The advantage of this deformation is that, in general, two different exponential scales~\eqref{Lambda-pm} emerge in the strong-coupling trans-series. The corresponding non-perturbative sectors satisfy certain relations that may break for special values of $a$, \cf~\eqref{gamma-ratio-Lambdapm}. Our goal is to present a similar generalization for the $\mathrm{O}(6)$ mass gap to obtain a complete trans-series and resurgence description for its strong-coupling expansion.

For simplicity, we start in medias res and define the tilted mass gap as
\begin{equation}\label{m-gen-fin}
    m(\alpha)=-{32g\cos(\alpha)\over \pi}e^{-4\pi g x_0^+}\left[f_0(-4\pi g x_0^+)U_0^-(4\pi g x_0^+)+f_1(-4\pi g x_0^+)U_1^-(4\pi g x_0^+)\right]\,,
\end{equation}
with $x_0^+=1/2-a$. It is straightforward to show that by substituting $a=1/4$, and using identities~\eqref{eq:U-V-rel} together with the quantization condition~\eqref{quant-fV}, this expression reduces to
\begin{equation}
    m\left(\alpha={\pi\over4}\right)\equiv m_{\mathrm{O}(6)}\,.
\end{equation}
Therefore, $m(\alpha)$ gives a one-parameter generalization of the $\mathrm{O}(6)$ mass gap.

It is important to emphasize that the definition in \eqref{m-gen-fin} is conjectured in analogy to the tilted cusp anomalous dimension, rather than derived from a deformation of the $\mathrm{O}(6)$ model. Its form is based on the fact that the physical mass gap can be expressed in terms of the special functions $f_n(x)$ and $U_n^\pm(x)$, which are related to the solution of the BES equation and evaluated at $x=-\pi g$, see (4.4) in~\cite{Basso:2009gh}. In the tilted case it is natural to enhance these special functions with their tilted generalizations. As discussed before, when computing non-perturbative contributions for the tilted cusp, one has to evaluate the special functions at $x=\mp 4\pi g x_j^\pm$, see the quantization conditions in Section \ref{Sec:Cusp-From-Bethe}. In the physical case $\alpha=\pi/4$, the zero closest to the origin is precisely $x=-\pi g$. Therefore, in the spirit of the tilted cusp anomalous dimension, it seems a natural generalization of the physical case to replace the argument $x=-\pi g$ in (4.4) of~\cite{Basso:2009gh} with the leading \emph{tilted} zero at $x=-4\pi g x_0^+$. Based on this we conjecture the form of \eqref{m-gen-fin}. Still, there is some freedom in adding terms to the right-hand side of \eqref{m-gen-fin} that vanish for $\alpha=\pi/4$ as well as in choosing the overall prefactor.
We used this freedom to choose the simplest possible modification.

An advantage of the definition~\eqref{m-gen-fin} is that it contains the same combination of functions $f_n(x)$ and $U^-_n(x)$ as the left-hand side of~\eqref{fU-ac}, evaluated at $x=2i\pi x_0^+$. Since for $0\leq \alpha\leq \pi/2$ and~$\Re x=\Re 2i\pi x_0^+=0$, we can use the representation in~\eqref{fU-ac}, to show that the tilted mass gap can also be expressed in terms of our well-known functions $a_\pm(ix)$ and $c_\pm$ as
\begin{equation}
\begin{aligned}\label{tilted-gap}
    m(\alpha)&={2(2\pi gx_0^+)^{a-1}\over\pi\sin(2\alpha)}e^{-4\pi g x_0^+}\times\\
    &\times\left[\left(g\partial_gc_+\right)a_-(-2\pi x_0^+)-c_+\left((3a-4\pi gx_0^+)a_-(-2\pi x_0^+)+g\partial_ga_-(-2\pi x_0^+)\right)+\right.\\
    &+\left(g\partial_gc_-\right)a_+(-2\pi x_0^+)-c_-\left((-a-4\pi gx_0^+)a_+(-2\pi x_0^+)+g\partial_ga_+(-2\pi x_0^+)\right)\left.\right] \,.
\end{aligned}
\end{equation}

Since, at strong coupling, the functions $a_\pm(ix)$ and $c_\pm$ are given by the trans-series~\eqref{apm-bpm} and~\eqref{cpm}, respectively, relation~\eqref{tilted-gap} implies that we can make a similar ansatz for the large-$g$ expansion of the tilted mass gap
\begin{equation}\label{m-strong-Lambda}
    m(\alpha)=M_0\sum_{n,m\geq0}\Lambda_-^{2n}\Lambda_+^{2m}m^{(n,m)}\,.
\end{equation}
The coefficient functions $m^{(n,m)}$ are given by a series in $1/g$. Here we also introduce the overall factor~$M_0$, which normalizes the series such that
\begin{equation}
    m^{(0,0)}=1+O\left(g^{-1}\right)\,.
\end{equation}
The advantage of relation \eqref{tilted-gap} is that by knowing the complete trans-series for $c_\pm$ and $a_\pm(ix)$ we can easily generate any corrections in~\eqref{m-strong-Lambda} to obtain a full strong-coupling expansion for the tilted mass gap. 

At this point, it is convenient to briefly recall the above discussion and summarize the method to compute the non-perturbative corrections for $m(\alpha)$. First, we have to determine the strong-coupling expansion of the functions $W_\pm$, $W_0$, $a_\pm(ix)$, and $b_\pm(ix)$, which was already done in previous work~\cite{Bajnok:2024bqr}. Inserting the solutions for $W_0$ and $W_\pm$ into the differential equations~\eqref{eq:DiffEq-c}, the non-perturbative coefficients of $c_\pm(g)$ can be uniquely determined. Finally, using~\eqref{tilted-gap}, any strong-coupling coefficient function of $m(\alpha)$ can be computed up to an arbitrary order in $1/g$.

Substituting~\eqref{apm-sol} and~\eqref{cp-sol} into~\eqref{tilted-gap}, we find the perturbative part of the tilted mass gap given by
\begin{equation}
\begin{aligned} \label{eq:pertTiltedMassGap}
    m^{(0,0)}&=1+\frac{a}{8\pi g}\left(\frac{2(1-a)}{1-2a}+\mathcal{I}_2\right)-\\
    &-\frac{a}{{2 (8\pi g)^2}} \left((1-a) \left(\frac{4 (2-a) (1-a)}{(1-2 a)^2}+\frac{4 (1-a) \mathcal{I}_2}{1-2 a}+\mathcal{I}_2^2\right)-(1-3 a) \mathcal{I}_3\right)+O\left(g^{-3}\right)\,,
\end{aligned}
\end{equation}
while the prefactor $M_0$ is chosen as
\begin{equation}\label{M0}
    M_0=\frac{4  \Gamma (1-a)}{\pi  \sin(2 \alpha)}\left(8 \pi  x_0^+ g\right)^a e^{-4 \pi  x_0^+ g}\,,
\end{equation}
and with $x_0^+=1/2-a$.

We can repeat the same procedure for non-perturbative corrections $m^{(n,m)}$. Using the trans-series results for $a_\pm(ix)$ and $c_\pm$, we find for the first few non-perturbative corrections
\begin{equation}
\begin{aligned} \label{eq:first-non-pert-massgap}
    m^{(1,0)}&=\frac{2^{1-2 a} (1-2 a)^{2 a-1} \Gamma (1-a)}{(8\pi g)\Gamma (a)}\left[1-\frac{2 (1-a) (2-3 a)+(1-2 a) (1-3 a) \mathcal{I}_2}{(1-2 a) (8\pi g)}+O\left(g^{-2}\right)\right]\,,\\
    m^{(2,0)}&=\frac{4^{1-2 a} (1-2 a)^{4 a-2} \Gamma (1-a)^2}{(8\pi g)^2\Gamma (a)^2}\left[1+O\left(g^{-1}\right)\right]\,,\\
    m^{(3,0)}&=O\left(g^{-3}\right)\,,\\
    m^{(0,1)}&=\frac{2^{2 a-1} (2 a-1) (2 a+1)^{-2 a} \Gamma (a+1)}{\Gamma (1-a)}\left[1-\frac{a (2 (1-a)+(1-2 a) \mathcal{I}_2)}{(1-2 a) (8\pi g)}+O\left(g^{-2}\right)\right]\,,\\
    m^{(1,1)}&=\frac{1}{4} (1-2 a)^{2 a+1} (2 a+1)^{1-2 a}\left[1-\frac{a (2 (a+1)-(1-2 a) \mathcal{I}_2)}{(1-2 a) (8\pi g)}+O\left(g^{-2}\right)\right]\,.
\end{aligned}
\end{equation}
In this way, we are able to compute corrections analytically up to $\Lambda_-^6$ and $\Lambda_+^6$ exponential orders and up to $O(g^{-2})$. Numerically, we have computed the first $20$ coefficients in $1/g$ up to $\Lambda_-^{16}\Lambda_+^{16}$ with $50$-digit precision. Moreover, focusing on the leading exponential orders, several hundred $1/g$ coefficients can be easily reached with even higher precision.

Interestingly, by computing further subleading corrections, we find that there is a natural generalization of equation (5.35) of~\cite{Dorigoni:2015dha} for arbitrary mixing angles $\alpha$, namely,
\begin{equation} \label{m-ratio-Lambdapm}
    {m^{(n+1,0)}\over m^{(n,0)}}={m^{(1,0)}\over m^{(0,0)}}\,,
\end{equation}
with $n\geq 0$. Notice that this relation is similar to the one obtained for the tilted cusp anomalous dimension in~\eqref{gamma-ratio-Lambdapm}. Furthermore, we can relate the perturbative part of the mass gap to the first non-perturbative correction of the cusp anomalous dimension through
\begin{equation}
    \Gamma^{(1,0)}=\frac{\left(x_0^+\right)^{2 a-1} }{8\pi g}{\Gamma (1-a)\over\Gamma (1+a)}\left[m^{(0,0)}\right]^2\,.
\end{equation}
The above relation is a generalization of~\eqref{cusp-mO6} for arbitrary mixing angles $\alpha$. The results above suggest a deep relation between the strong-coupling trans-series of the tilted cusp anomalous dimension and the tilted mass gap. We will discuss this in more detail in Section~\ref{Sec:Tilted-mass-gap}.

\subsection{The $\mathrm{O}(6)$ mass gap ($a=1/4$)} \label{Sec:O6-mass-gap-res}

Having worked out the $\mathrm{O}(6)$ \emph{tilted} mass gap, we would now like to take the special value $a=1/4$ to compare the results for the leading large-$g$ expansion with~\cite{Basso:2009gh}. For the perturbative part, the expression~\eqref{eq:pertTiltedMassGap} becomes
\begin{equation}\label{m-pert-spec}
    m\left(\alpha={\pi\over 4}\right)=\frac{\sqrt{2} (2\pi g)^{1\over 4} e^{-\pi  g}}{\Gamma \left(\frac{5}{4}\right)}\left[1+\frac{3-6 \log 2}{4(8 \pi  g)}+\frac{16 K-63+108\log2-108(\log2)^2}{32(8 \pi  g)^2}+O\left(g^{-3}\right)\right]\,.
\end{equation}
Here, we also substituted the first moments with $\mathcal I_2=-{6\log 2}$ and $\mathcal I_3={16K}$, with $K$ being the Catalan constant.
Beautifully, the first three orders perfectly match (4.12) of~\cite{Basso:2009gh}.

The mass gap, including non-perturbative corrections, can then be obtained by substituting~\eqref{eq:first-non-pert-massgap} into the trans-series~\eqref{m-strong-Lambda} and setting the parameter $a=1/4$. Writing everything in terms of the effective parameter $\Lambda$ introduced in~\eqref{eq:EffectiveLambda} and including up to three non-perturbative orders, we obtain
\begin{equation}\label{m-np-spec}
\begin{aligned}
    m_{\mathrm{O}(6)}=\frac{\sqrt{2} (2\pi g)^{1\over 4} e^{-\pi  g}}{\Gamma \left(\frac{5}{4}\right)}&\left[\left(1+\frac{3-6 \log 2}{4 (8\pi  g)}+\frac{16 K-63+108\log2-108(\log2)^2}{32 (8\pi  g)^2}+O\left(g^{-3}\right)\right)-\right.\\
    -&{\Lambda^2\over 8\pi g}\left(1-\frac{15-6\log 2}{4 (8 \pi  g)}-\frac{208 K-1161+900\log2-180(\log2)^2}{32 (8\pi  g)^2}+O\left(g^{-3}\right)\right)+\\
    +&{\Lambda^4\over(8\pi g)^2}\left(1-\frac{33-18 \log 2}{4 (8\pi  g)}+O\left(g^{-2}\right)\right)-\\
    -&\frac{\Lambda^6}{(8\pi g)^2} \frac{8 \sqrt{\frac{2}{3}} \Gamma \left(\frac{5}{4}\right)^5}{\pi  \Gamma \left(\frac{3}{4}\right)^3}\left(1-\frac{3-6\log 2-{\sqrt{3}\Gamma\left(3\over4\right)^4\Gamma\left(5\over4\right)^{-4}}}{4(8\pi g)}+O\left(g^{-2}\right)\right)-\\
    -&\left. {\Lambda^8\over(8\pi g)^2}\frac{2 \sqrt{6} \Gamma \left(\frac{5}{4}\right)^5}{\pi  \Gamma \left(\frac{3}{4}\right)^3}\left(1-\frac{5+6\log 2}{4(8\pi g)}+O\left(g^{-2}\right)\right)+\dots\right]\,.
\end{aligned}
\end{equation}
Again, this is in perfect agreement\footnote{Note that here we corrected a typo in the Stokes constant at $\Lambda^6$ presented in eq. (C.13) of~\cite{Dorigoni:2015dha}.} with the known results in the literature~\cite{Basso:2009gh,Dorigoni:2015dha}.

Notice that relation~\eqref{m-ratio-Lambdapm} holds for the physical value $a=1/4$ as well and relates non-perturbative corrections with different powers of $\Lambda_-$. In~\cite{Dorigoni:2015dha}, the authors conjectured a general rule for the non-perturbative sectors of the mass gap expanded in powers of $\Lambda$, see equation (5.35) of~\cite{Dorigoni:2015dha}. However, they observed that their rule for the ratios breaks down at order $\Lambda^6$. The reason for this is exactly the mixing of the sectors $\Lambda_-$ and $\Lambda_+$ first appearing at order $\Lambda^6$, \cf~\eqref{scale-mixture}. A huge advantage of our method of introducing the tilted mass gap is that we can move slightly away from the physical value $a=1/4$ to separate the different exponential scales. This reveals properties of the strong-coupling structure that are otherwise hidden for the physical case. 

\section{Universal strong-coupling structure} \label{Sec:Universal-Structure}

The exact observables studied in this paper admit a trans-series expansion at strong coupling. The series in $1/g$ for each (non-)perturbative sector are asymptotic and give rise to a rich analytic structure in the Borel plane.
Resurgence provides a framework that unifies these perturbative and non-perturbative contributions into a single trans-series description and makes the relations between them precise, \cf~\cite{Marino:2012zq,Dorigoni:2014hea,Aniceto:2018bis} for an overview. In particular, the singularity structure of the Borel transform and the associated Alien derivatives connect different sectors of the trans-series. In this Section we analyze the resurgent structure of the Fredholm determinant observables, the tilted cusp anomalous dimension, and the tilted mass gap. We will use the fact that these quantities exhibit a remarkably universal strong-coupling structure~\cite{Bajnok:2025lji,Boldis:2026rkb,Bajnok:2026xri}, allowing their perturbative expansions, non-perturbative corrections, and Stokes phenomena to be described within a common resurgent framework.

\subsection{Determinant} \label{Sec:Universal-Resurgence-Determinant}

So far we have investigated the connection between the tilted cusp anomalous dimension and the tilted mass gap. We found that both quantities can be written in terms of the same building blocks $a_\pm(ix)$, $b_\pm(ix)$, and $c_\pm$, which can be obtained from a set of differential equations. In this Section we investigate the universal resurgence structure for the tilted mass gap at strong coupling. 
In order to achieve this, we will build on previous results for the determinant observables. We found in~\cite{Bajnok:2025lji} that these observables possess beautiful and universal resurgence relations. 
This was then generalized in~\cite{Boldis:2026rkb} for the tilted cusp anomalous dimension, and the cusp itself ($a=1/4$) was considered in~\cite{Bajnok:2026xri}. The goal is therefore to relate the mass gap to determinant observables and investigate the resulting resurgence behavior.

Before turning to the tilted mass gap, let us recall the results for determinant observables $Z_\ell(g|\alpha)$, which are defined as in~\eqref{F-def}. Using a notation similar to~\cite{Boldis:2026rkb}, we can write the trans-series as
\begin{equation}\label{trans-D}
    Z_\ell(\alpha)=A_\ell \sum_{\delta^+,\delta^-}(8\pi g)^{2a\Delta}e^{i\alpha\Delta }e^{-8\pi g \left(\sum_{l\in \delta^+}  x_l^++\sum_{j\in \delta^- } x_j^-\right)}Z_\ell^{(\delta^+,\delta^-)}\,.
\end{equation}
The summation goes over all pairs of unordered multisets $\delta^+,\delta^-$ of non-negative integers, possibly containing the same multiple times. 
%, which ensures that the trans-series is at most first order in every exponential weight $e^{-8\pi g x_l^\pm}$, where 
Again, $x_l^\pm$ denotes the zeros of the modified symbol $\chi_\alpha(x)$, and their values are given in \eqref{zeros}. These weights correspond to the powers of the trans-series parameters, and the exponent $x_l^\pm$ is related to the zeros of the functions $\Phi_\pm(x)$. Moreover, we define $\Delta$ as the difference in cardinality of the two sets for a fixed pair of $\delta^+$ and $\delta^-$ as $\Delta=|\delta^+|-|\delta^-|$.

The reader might wonder why the trans-series~\eqref{trans-D} is written in the basis $(\delta^+,\delta^-)$ instead of the basis $(n,m)$ used in~\eqref{Z-strong-Lambda}. The advantages of the new choice will become clear in the following. However, we already stress at this point that this notation is the natural basis for the general resurgence structure observed and investigated in~\cite{Bajnok:2025mxi,Boldis:2026rkb,Bajnok:2026xri}.
The trans-series parameters $\Lambda_\pm$ scale as
\begin{equation}
    \Lambda_-^2\sim e^{-8\pi g x_0^+}\,,\qquad \Lambda_+^2\sim e^{-8\pi g x_0^-}\,.
\end{equation}
Combinations of higher powers of $\Lambda_-$ and $\Lambda_+$ can also be a mixture of weights $e^{-8\pi gx_j^\pm}$ and their prefactors $(8\pi g)^{2a\Delta}e^{i\alpha\Delta}$. However, the non-perturbative corrections $Z_\ell^{(n,m)}$ and $Z_\ell^{(\delta^+,\delta^-)}$ in the two different notation are not in a one-to-one correspondence. 
Therefore, different terms in~\eqref{trans-D} could contribute at the same orders of $\Lambda_-$ and $\Lambda_+$ in~\eqref{Z-strong-Lambda}, for example
\begin{equation}
\begin{aligned}\label{Z-rel}
    &Z_\ell^{(0,0)}=Z^{(\{\},\{\})}\,,
    &&Z_\ell^{(1,1)}=Z_\ell^{(\{0\},\{0\})}\,,\\
    &Z_\ell^{(1,0)}=Z_\ell^{(\{0\},\{\})}\,,
    &&Z_\ell^{(0,1)}=Z_\ell^{(\{\},\{0\})}\,,\\
    &Z_\ell^{(2,0)}=Z_\ell^{(\{0,0\},\{\})}\,,
    &&Z_\ell^{(0,2)}=Z_\ell^{(\{\},\{0,0\})}\,,\\
    &Z_\ell^{(3,0)}=Z_\ell^{(\{0,0,0\},\{\})}\,,
    &&Z_\ell^{(0,3)}=Z_\ell^{(\{\},\{0,0,0\})}\,,\\
    &Z_\ell^{(2,1)}=Z_\ell^{(\{1\},\{\})}+Z_\ell^{(\{0,0\},\{0\})}\,,
    &&Z_\ell^{(1,2)}=Z_\ell^{(\{\},\{1\})}+Z_\ell^{(\{0\},\{0,0\})}\,,\\
    &Z_\ell^{(3,2)}=Z_\ell^{(\{2\},\{\})}+Z_\ell^{(\{0,0\},\{1\})}+Z_\ell^{(\{0,1\},\{0\})}+Z_\ell^{(\{0,0,0\},\{0,0\})}\,,
    &&\dots
\end{aligned}
\end{equation}
A further discussion and more examples on this structure can be found in~\cite{Boldis:2026rkb}.

Further, in~\eqref{trans-D} we have the trans-series coefficient function $Z_\ell^{(\delta^+,\delta^-)}$ that we write as 
\begin{equation}\label{Z-np}
    Z_\ell^{(\delta^+,\delta^-)}=(8\pi g)^{-\Delta^2}S_\ell^{(\delta^+,\delta^-)}\mathcal{D}_\ell^{(\delta^+,\delta^-)}\,,
\end{equation}
with the Stokes constant $S_\ell^{(\delta^+,\delta^-)}$ and the functions $\mathcal{D}^{(\delta^+,\delta^-)}(g)$. The latter are once again given by a series in $1/g$.
Finally, the overall factor in~\eqref{trans-D} is chosen in such a way that the perturbative term is normalized as $S^{(\{\},\{\})}=1$. In general, the normalization in~\eqref{Z-np} is chosen as $\mathcal{D}^{(\delta^+,\delta^-)}(g)=1+O(1/g)$.

As mentioned, the non-perturbative corrections are related to the perturbative series through a simple rule~\cite{Boldis:2026rkb}. Let us denote the perturbative series as
\begin{equation} \label{eq:Def-D_ell}
    Z_\ell^{(\{\},\{\})}=\mathcal{D}_\ell^{(\{\},\{\})}\equiv \mathcal{D}_\ell\left[a,\,\mathcal{I}_n\right]\,.
\end{equation}
This function coincides with the expansion in~\eqref{eq:pertDet} and~\eqref{eq:pertDetCoeff} and we have highlighted the dependence on the parameter $a$ and the moments $\mathcal{I}_n$. Here, the dependence on $a$ only represents its {\it explicit} dependence, and the moments $\mathcal{I}_n$ are treated as independent symbolic quantities.

Importantly, the non-perturbative contributions $\mathcal{D}^{(\delta^+,\delta^-)}$ are then given by the \emph{shift relations}, where the parameter $a$ and the moments are shifted simultaneously in the following way
\begin{equation}\label{D-rule}
    \mathcal{D}_\ell^{(\delta^+,\delta^-)}=\mathcal{D}_\ell\left[a-\Delta,\,\mathcal{I}^{(\delta^+,\delta^-)}_n\right]\,.
\end{equation}
The shifted moments $\mathcal{I}^{(\delta^+,\delta^-)}_n$ are given by
\begin{equation} \label{eq:Def-shifted-moment}
\mathcal{I}_n^{(\delta^+,\delta^-)}=\mathcal{I}_n-(-1)^{n-1}\sum_{l\in\delta^+}\frac{2}{(x_l^+)^{n-1}}+\sum_{j\in\delta^-}\frac{2}{(x_j^-)^{n-1}}\,.
\end{equation}
The shift $a\to a-\Delta$ in~\eqref{D-rule} refers only to the explicit dependence on the parameters $a$ and $\mathcal{I}_n$, while the zeros $x_j^\pm$ are treated as \emph{independent} quantities. These relations highlight why it is necessary to abandon the notation~\eqref{Z-strong-Lambda} and instead write the trans-series in the form~\eqref{trans-D}. Evidently these shifts are naturally written in terms of the zeros $x_j^\pm$ of the function $\Phi_\pm(x)$ as opposed to the counters of the non-perturbative order.

Furthermore, there are \emph{recurrence relations} for the Stokes constants depending on the respective non-perturbative direction. For instance, adding the $k$-th non-perturbative correction to the first set of ${(\delta^{+},\delta^{-})}$ we have \cite{Boldis:2026rkb}
\begin{equation}
\begin{aligned}\label{Sp-fin}
S_\ell^{(\delta^{+}\cup\{k\},\delta^-)}=&-(-1)^{\ell+\Delta}\frac{\Gamma\left(\ell-a+\Delta+1\right)}{\Gamma\left(\ell+a-\Delta\right)}\frac{\Gamma^2\left(k+\frac{1}{2}-a\right)}{\Gamma^2\left(\frac{1}{2}-a\right)\Gamma^2\left(k+1\right)}\times\\
&\times\frac{\prod_{j\in \delta^+}\left(\frac{j-k}{j+\frac{1}{2}-a}\right)^{2}}{\prod_{j\in\delta^-}\left(\frac{j+k+1}{j+\frac{1}{2}+a}\right)^{2}}\left(k+\frac{1}{2}-a\right)^{2a-1-2\Delta}S_\ell^{(\delta^{+},\delta^{-})}\,,\\
\end{aligned}
\end{equation}
whereas adding the $k$-th non-perturbative correction to the second set of ${(\delta^{+},\delta^{-})}$ leads to
\begin{equation}
\begin{aligned} \label{Sm-fin}
S_\ell^{(\delta^{+},\delta^{-}\cup\{k\})}=&-(-1)^{\ell-\Delta}\frac{\Gamma\left(\ell+a-\Delta+1\right)}{\Gamma\left(\ell-a+\Delta\right)}\frac{\Gamma^2\left(k+\frac{1}{2}+a\right)}{\Gamma^2\left(\frac{1}{2}+a\right)\Gamma^2\left(k+1\right)}\times\\
&\times\frac{\prod_{j\in \delta_j^-}\left(\frac{j-k}{j+\frac{1}{2}+a}\right)^{2}}{\prod_{j\in \delta_j^+}\left(\frac{j+k+1}{j+\frac{1}{2}-a}\right)^{2}}\left(k+\frac{1}{2}+a\right)^{-2a-1+2\Delta}S_\ell^{(\delta^{+},\delta^{-})}\,.
\end{aligned}
\end{equation}
Notice that the Stokes constants vanish if $k$ is added to a set $\delta^{\pm}$ that already contains $k$. This is due to the product in the numerator in the respective second line of~\eqref{Sp-fin},~\eqref{Sm-fin}, as it contains a factor $(j-k)$ with $j\in \delta^\pm$. This has an important consequence regarding the trans-series~\eqref{trans-D}, namely this property ensures that the strong-coupling expansion is at most first order in every exponential weight $e^{-8\pi g x_l^\pm}$, and hence many terms in~\eqref{Z-rel} are, in fact, equal to zero. Finally, it is also important to note that all the Stokes constants $S^{(\delta^+,\delta^-)}$ and the $1/g$ coefficients in $Z_\ell^{(\delta^+,\delta^-)}$ are real, hence the non-perturbative sectors obtain complex value only due to the multiplicative factors $e^{i\alpha\Delta}$ at each exponential orders.

Having collected all these ingredients, the full trans-series for the determinant can be generated solely from the perturbative part. As an instructive example, we present the form of the $Z_0^{(\{i\},\{\})}$ correction. According to~\eqref{Z-np} and~\eqref{D-rule}, its $1/g$ expansion takes a similar form as~\eqref{eq:pertDet} and reads
\begin{equation}\label{Z10}
    Z_0^{(\{i\},\{\})}={S_0^{(\{i\},\{\})}\over 8\pi g}\left[ 1 + (a-1)^2 \sum_{k \ge 1} \frac{(-1)^k}{k!} \frac{f_k^{(\{i\},\{\})}}{(8\pi g)^k}\right]\,.
\end{equation}
The Stokes constant $S_0^{(\{i\},\{\})}$ takes the form of
\begin{equation}\label{Stokes-eg}
S_0^{(\{i\},\{\})}=-\frac{\Gamma\left(1-a\right)}{\Gamma\left(a\right)}\frac{\Gamma^2\left(i+\frac{1}{2}-a\right)}{\Gamma^2\left(\frac{1}{2}-a\right)\Gamma^2\left(i+1\right)}\left(i+\frac{1}{2}-a\right)^{2a-1}\,,
\end{equation}
and the first coefficients are given by
\begin{equation}
\begin{aligned}\label{f10}
    f^{(\{i\},\{\})}_1&=\mathcal{\mathcal{I}}_2+2(1/2+i-a)^{-1}\,,\\
    f^{(\{i\},\{\})}_2&=((a-1)^2+1)\left(\mathcal{I}_2+2(1/2+i-a)^{-1}\right)^2+2(a-1)\left(\mathcal{I}_3-2(1/2+i-a)^{-2}\right)\,,\\
    f^{(\{i\},\{\})}_3&=\left((a-1)^2+1\right) \left((a-1)^2+2\right)\left(\mathcal{I}_2+2(1/2+i-a)^{-1}\right)^3+\\
    &+6 (a-1) \left((a-1)^2+2\right)\left(\mathcal{I}_2+2(1/2+i-a)^{-1}\right)\left(\mathcal{I}_3-2(1/2+i-a)^{-2}\right)+\\
    &+2(5 (a-1)^2+1)\left(\mathcal{I}_4+2(1/2+i-a)^{-3}\right)\,.
\end{aligned}
\end{equation}
These expressions were obtained from~\eqref{eq:pertDetCoeff} by setting $\ell=0$, shifting the explicit $a$ dependence by $a\to a-1$ and substituting $\mathcal{I}_n\to\mathcal I_n-(-1)^{n-1}2/(x_i^+)^{n-1}$.

Moving to higher orders in the $1/g$ expansion, the coefficients $f_k$ in~\eqref{Z10} start to grow factorially, indicating that the series $Z^{(\delta^+,\delta^-)}$ is asymptotic. Therefore, a careful resurgence analysis is necessary before the trans-series~\eqref{trans-D} can be resummed. The above results can be summarized as follows: to resum the asymptotic series in \eqref{trans-D}, one first has to take the Borel transform of each $1/g$ series $Z_\ell^{(\delta_ +,\delta_-)}$. Investigating the analytic structure in the Borel plane, we find that $Z_\ell^{(\delta_ +,\delta_-)}$ has Borel singularities at $s=x_l^\pm$ (and at their linear combinations) with $x_l^\pm$ such that their labels are not included in the sets $\delta_\pm$. When going back from the Borel plane to the physical plane, the trans-series structure  \eqref{trans-D} ensures the cancellation of the ambiguities arising from lateral Borel resummation.

Going to higher non-perturbative orders, some of the singularities on the positive real axis of the Borel plane, which are related to the zeros $x_j^\pm$ of $\Phi_\pm(x)$, become singularities on the negative real line, which are related to the poles $y_j$ of $\Phi_\pm(x)$. The new $1/g$ series at the given non-perturbative order looks exactly as the original perturbative sector, but now the zero $x_j^\pm$ plays the role of an additional pole $y_j$. Hence, the corresponding moment changes as in~\eqref{eq:Def-shifted-moment} and accordingly the Stokes constants as in~\eqref{Sp-fin} and~\eqref{Sm-fin}.

A more detailed discussion of the resurgence structure of determinant observables can be found in~\cite{Boldis:2026rkb,Bajnok:2026xri}. It is important to note that the trans-series in~\eqref{trans-D} corresponds to the lateral resummation~$\mathcal S_+$ in the upper half of the Borel plane.

All the resurgence relations following from the structure discussed above can be elegantly recast in terms of the Stokes automorphism, which encodes the difference between the lateral resummations of the trans-series. Its action is generated by the so-called \emph{Alien derivatives} $\Delta^\pm$. These operators relate various non-perturbative sectors. For each exponential scale $e^{-8\pi g x_j^\pm}$ we associate an Alien derivative $\Delta^\pm_j$. As was shown in~\cite{Boldis:2026rkb}, these Alien derivatives act on the different non-perturbative sectors as
\begin{equation}
\begin{aligned}\label{AD-pm}
    &\Delta^+_j Z_\ell^{(\delta^+,\delta^-)}=-2i\sin\left(\alpha\right)\,(8\pi g)^{2a}Z_\ell^{(\delta^+\cup\{j\},\delta^-)}\,,\\
    &\Delta^-_j Z_\ell^{(\delta^+,\delta^-)}= 2i\sin\left(\alpha\right)\,(8\pi g)^{-2a}Z_\ell^{(\delta^+,\delta^-\cup\{j\})} \,.
\end{aligned}
\end{equation}
Furthermore, the non-perturbative correction $Z_\ell^{(\delta^+,\delta^-)}$ is independent of the order in which the Alien derivatives are applied to obtain it from the perturbative part $Z_\ell^{(\{\},\{\})}$. In particular, this means
\begin{equation}\label{AD-pm-algebra}
 \left[\Delta^+_l,\Delta^+_j\right]=\left[\Delta^-_l,\Delta^-_j\right]=\left[\Delta^+_l,\Delta^-_j\right]=0\,,
\end{equation}
for all integers $j,l\geq 0$. 

Moreover, the action of the Alien derivatives on $Z_\ell^{(\delta^+,\delta^-)}$ is nilpotent
\begin{equation}\label{AD-sqr}
\left(\Delta^\pm_j\right)^2 Z_\ell^{(\delta^+,\delta^-)}=0\,,
\end{equation}
\ie~acting on $Z_\ell^{(\delta^+,\delta^-)}$ with $\Delta_j^\pm$ where $j\in\delta^+$ or $j\in \delta^-$ gives zero.
Hence, non-perturbative corrections that are at least second order in one of the exponential weights $e^{-8\pi g x_l^\pm}$ vanish. Notice that this is also ensured by the vanishing of the Stokes constants~\eqref{Sp-fin},~\eqref{Sm-fin} for such corrections.

It is important to note that the factors $\sin(\alpha)$ on the right-hand side of~\eqref{AD-pm} are related to the phase~$e^{i\alpha\Delta}$ appearing in the trans-series~\eqref{trans-D}. More precisely, each exponential correction~$e^{i\alpha\Delta}Z_\ell^{(\delta^+,\delta^-)}$ is complex valued. The Alien algebra only fixes the imaginary part
while the real part in general is not related to the Alien derivatives of the preceding non-perturbative sectors. However in our exceptional situation, all the $1/g$ corrections and the Stokes constants $S_\ell^{(\delta^+,\delta^-)}$ in~$Z_\ell^{(\delta^+,\delta^-)}$ are real, and the only ambiguity comes from the overall phase~$e^{i\alpha\Delta}$. Therefore up to the overall factor, which comes from taking the imaginary part~$e^{i\alpha\Delta}$, non-perturbative corrections are directly given by the Alien derivatives of the preceding sectors, hence all corrections can be completely generated by the Alien algebra itself.

\subsection{Tilted cusp}\label{cusp-sec}

The strong-coupling structure of the tilted cusp was investigated in~\cite{Boldis:2026rkb}. In this Subsection we will collect some important results and expand on the Alien algebra.

Equipped with the previous results, we can now investigate the tilted cusp, which can be expressed as a ratio of two determinants, see~\eqref{tilted-cusp-det}. Hence, we can parametrize the strong-coupling expansion as
\begin{equation}\label{trans-cusp}
    \Gamma_{\mathrm{cusp}}(\alpha)= \frac{8g a}{\sin (2\alpha)}\sum_{ \delta^+, \delta^-}(8\pi g)^{2a \Delta}e^{i\alpha \Delta }e^{-8\pi g \left(\sum_{l\in \delta^+}  x_l^++\sum_{j\in  \delta^- } x_j^-\right)}\Gamma^{( \delta^+, \delta^-)}\,.
\end{equation}
Here $\delta^\pm$ are again multisets of integer numbers. The series is normalized in such a way that
\begin{equation}
    \Gamma^{(\{\},\{\})}=1+\mathcal{O}\left({1\over g}\right)\,.
\end{equation}
As in the previous subsection, this trans-series is a restructured version of~\eqref{cusp-strong-Lambda}.

Considering the perturbative order of the tilted cusp~\eqref{tilted-cusp-det}, we can write it as the ratio of the perturbative parts of the respective determinants
\begin{equation}\label{gamma-pert}
    \Gamma^{(\{\},\{\})}={Z_1^{(\{\},\{\})}\over Z_0^{(\{\},\{\})}}\,.
\end{equation}

In the following, we investigate the strong-coupling structure of the tilted cusp anomalous dimension in terms of the same Alien derivatives that were used to describe the strong-coupling structure of the determinants $Z_\ell(\alpha)$. This is motivated by the following two properties:
\begin{enumerate}
    \item The cusp anomalous dimension can be expressed as the ratio of two determinants, which means that each non-perturbative sector in \eqref{trans-cusp} can be written as a function of different sectors of $Z_0(\alpha)$ and $Z_1(\alpha)$,
    \item In certain situations Alien derivatives behave similarly to ordinary derivatives, e.g. they satisfy the Leibniz rule when acting on products of different non-perturbative objects.
\end{enumerate}

We claim that up to a slight modification, the Alien algebra in~\eqref{AD-pm} holds for the tilted cusp as well, explicitly
\begin{equation}
\begin{aligned}\label{AD-gamma}
    &\Delta^+_j\Gamma^{(\delta^+,\delta^-)}=-{2i\sin\left(\alpha\right)}(N^+_j+1)\,(8\pi g)^{2a}\Gamma^{(\delta^+\cup\{j\},\delta^-)}\,,\\
    &\Delta^-_j\Gamma^{(\delta^+,\delta^-)}= {2i\sin\left(\alpha\right)}(N^-_j+1)\,(8\pi g)^{-2a}\Gamma^{(\delta^+,\delta^-\cup\{j\})}\,.
\end{aligned}
\end{equation}
Again, by acting with the Alien derivative $\Delta^\pm_j$ the element $j$ is added to the array $\delta^\pm$. Compared to~\eqref{AD-pm} we see an additional factor containing $(N^\pm_j+1)$, where the numbers $N^\pm$ count the number of $j$'s in the sets $\delta^\pm$. Hence, for example, we have
\begin{equation}
\begin{aligned}
    &\Delta^+_0\Gamma^{(\{\},\{\})}=-2i\sin(\alpha)(8\pi g)^{2a}\Gamma^{(\{0\},\{\})}\,,\\
    &\Delta^+_0\Gamma^{(\{0\},\{\})}=-4i\sin(\alpha)(8\pi g)^{2a}\Gamma^{(\{0,0\},\{\})}\,,\\
    &\Delta^+_0\Gamma^{(\{0,0\},\{\})}=-6i\sin(\alpha)(8\pi g)^{2a}\Gamma^{(\{0,0,0\},\{\})}\,,\\
    &\Delta^-_0\Gamma^{(\{0,0,0\},\{\})}=2i\sin(\alpha)(8\pi g)^{-2a}\Gamma^{(\{0,0,0\},\{0\})}\,.
\end{aligned}
\end{equation}

Importantly, the action of the Alien derivatives on the non-perturbative corrections of the cusp anomalous dimension in~\eqref{AD-gamma} is the generalization of the action in~\eqref{AD-pm}. In fact, applying~\eqref{AD-gamma} to the determinant reproduces~\eqref{AD-pm}, since every index can appear at most once in $\delta^{\pm}$ due to~\eqref{AD-sqr}. Put differently, the strong-coupling expansion of the determinant is at most first order in every exponential weight $e^{-8\pi gx_l^\pm}$. However, for the tilted cusp, higher orders of the exponential weight can appear due to the ratio of determinants. 
Ultimately, the prefactors $(N^\pm_j+1)$ contribute only for higher order derivatives, while the first-order Alien derivatives~\eqref{AD-gamma} coincide with~\eqref{AD-pm}. 
Consequently, the relation~\eqref{AD-sqr} is only valid for determinants. The factors $\sin(\alpha)$ in~\eqref{AD-gamma} are again related to the fact that the Alien algebra only fixes the imaginary part of the subleading contributions. However, the coefficients in~$\Gamma^{(\delta^+,\delta^-)}$ are real, and the only ambiguity comes from the phase~$e^{i\alpha\Delta}$, which means that it is possible to effectively generate all non-perturbative corrections using the Alien algebra~\eqref{AD-gamma}.

Now, let us use the Alien algebra assuming that the Alien derivative $\Delta^\pm$ acts like an ordinary \emph{differentiation}. Starting from the perturbative part~\eqref{gamma-pert} we take its first Alien derivative as
\begin{equation}
\begin{aligned}
    \label{first-AD}
    \Delta_0^+\Gamma^{(\{\},\{\})} &=\Delta_0^+ {Z_1^{(\{\},\{\})}\over Z_0^{(\{\},\{\})}} \\
    &={\left[\Delta^+_0Z_1^{(\{\},\{\})}\right]Z_0^{(\{\},\{\})}-\left[\Delta^+_0Z_0^{(\{\},\{\})}\right]Z_1^{(\{\},\{\})}\over  \left[Z_0^{(\{\},\{\})}\right]^2} \\
&=-2i\sin(\alpha)(8\pi g)^{2a}{Z_1^{(\{0\},\{\})}Z_0^{(\{\},\{\})}-Z_0^{(\{0\},\{\})}Z_1^{(\{\},\{\})}\over  \left[Z_0^{(\{\},\{\})}\right]^2}\,,
\end{aligned}
\end{equation}
where we used~\eqref{AD-pm} to go from the first line to the second. Using~\eqref{AD-gamma} we can read off
\begin{equation}\label{gamma-lead}
    \Gamma^{(\{0\},\{\})}={Z_1^{(\{0\},\{\})}Z_0^{(\{\},\{\})}-Z_0^{(\{0\},\{\})}Z_1^{(\{\},\{\})}\over  \left[Z_0^{(\{\},\{\})}\right]^2}\,.
\end{equation}
Remarkably, by applying the Alien derivatives the non-perturbative corrections to the tilted cusp can be expressed in terms of the determinants, and the result coincides with the non-perturbative sector obtained by expanding the ratio~\eqref{tilted-cusp-det} for small $e^{-8\pi g x_l^\pm}$.
Moreover, using~\eqref{D-rule} and~\eqref{Sp-fin},~\eqref{Sm-fin} we can relate the expression to the perturbative quantity $Z_\ell^{(\{\},\{\})}$.
It can be explicitly checked that~\eqref{gamma-lead} perfectly reproduces the leading non-perturbative contribution of the tilted cusp.

As another example, let us repeatedly act with the Alien derivative $\Delta^+_0$ on~\eqref{gamma-lead} to find the subleading corrections $\Gamma^{(\{0,0\},\{\})}$ and $\Gamma^{(\{0,0,0\},\{\})}$, which are then given by
\begin{equation}
\begin{aligned}\label{ind-init}
    &\Gamma^{(\{0,0\},\{\})}=\Gamma^{(\{0^{(2)}\},\{\})}=-{\Delta^+_0\Gamma^{(\{0\},\{\})}\over4i(8\pi g)^{2a}\sin(\alpha)}=-{Z_0^{(\{0\},\{\})}\over Z_0^{(\{\},\{\})}}\Gamma^{(\{0\},\{\})}\,,\\
    &\Gamma^{(\{0,0,0\},\{\})}=\Gamma^{(\{0^{(3)}\},\{\})}=-{\Delta^+_0\Gamma^{(\{0,0\},\{\})}\over 6i(8\pi g)^{2a}\sin(\alpha)}=-{Z_0^{(\{0\},\{\})}\over Z_0^{(\{\},\{\})}}\Gamma^{(\{0,0\},\{\})}\,.
\end{aligned}
\end{equation}
Note that by abuse of notation in this context we understand $\{0^{(m)}\}$ as the set that contains $m$-times of the entry $0$.
These relations were also tested both analytically and numerically. 

We compared non-perturbative corrections in~\eqref{trans-cusp} obtained by the Alien algebra~\eqref{AD-gamma} with corrections in the $(n,m)$ notation in~\eqref{cusp-strong-Lambda}. We used numerical data for $a=1/4$ and $a=1/(2\sqrt{2})$ up to $O(g^{-20})$ and analytical expressions for arbitrary $a$ up to $O(g^{-5})$ with exponential orders up to $\Lambda_-^6$ and $\Lambda_+^6$. We found perfect agreement between the two notations, which suggests that our conjecture~\eqref{AD-gamma} and the underlying assumptions are indeed valid and can be used to generate arbitrary non-perturbative corrections to the tilted cusp anomalous dimension.

Moreover, we can prove that the following ratios hold
\begin{equation}\label{gamma-rat-01}
    {\Gamma^{(\{0^{(n+1)}\},\{\})}\over \Gamma^{(\{0^{(n)}\},\{\})}}=-{Z_0^{(\{0\},\{\})}\over Z_0^{(\{\},\{\})}}\,.
\end{equation}
This relation follows from the action~\eqref{AD-gamma} and holds for $n\geq 1$. This relation is equivalent to~\eqref{gamma-ratio-Lambdapm}. In Appendix~\ref{App:Proof} we give a detailed proof for~\eqref{gamma-rat-01}, based on the assumptions that the Alien algebra in \eqref{AD-gamma} holds, and the Alien derivatives behave as ordinary differentiation when acting on products and ratios of non-perturbative sectors $Z_{\ell}^{(\delta^+,\delta^-)}$ of the determinant. 

Finally, using the Alien algebra we can generalize~\eqref{gamma-ratio-Lambdapm} and~\eqref{ind-init} even further. Considering the ratio of the $(n+1)$-th and $n$-th Alien derivative $(\Delta_i^\pm)$ we find
\begin{equation}\label{gamma-ratio}
    {\Gamma^{(\{i^{(n+1)}\},\{\})}\over \Gamma^{(\{i^{(n)}\},\{\})}}= -{Z_0^{(\{i\},\{\})}\over Z_0^{(\{\},\{\})}} \,, 
    \qquad {\Gamma^{(\{\},\{i^{(n+1)}\})}\over \Gamma^{(\{\},\{i^{(n)}\})}}= -{Z_0^{(\{\},\{i\})}\over Z_0^{(\{\},\{\})}}\,.
\end{equation}
For $i=0$ this perfectly reproduces~\eqref{gamma-ratio-Lambdapm} and for $n=1$ and $n=2$ we already presented an explicit check in~\eqref{ind-init}. Once again we refer to Appendix~\ref{App:Proof} for the full proof. These relations encode a general relation between certain non-perturbative sectors. 

It is important to note that we \emph{only} found a simple relation between non-perturbative corrections for the ones presented in~\eqref{gamma-ratio}.

\subsection{Tilted mass gap} \label{Sec:Tilted-mass-gap}

Finally, we turn to the strong-coupling structure of the tilted mass gap. Similarly to the trans-series for the tilted cusp anomalous dimension~\eqref{trans-cusp}, the mass gap can be parametrized as 
\begin{equation}\label{m-trans-xpm}
    m(\alpha)=M_0\sum_{ \delta^+, \delta^-}(8\pi g)^{2a \Delta}e^{i\alpha \Delta }e^{-8\pi g \left(\sum_{l\in  \delta^+}  x_l^++\sum_{j\in  \delta^- } x_j^-\right)}m^{( \delta^+, \delta^-)}\,.
\end{equation}
Choosing once again the normalization of the perturbative part so that 
\begin{equation}\label{M-norm}
    m^{(\{\},\{\})}=1+\mathcal{O}\left({1\over g}\right)\,,
\end{equation}
and the factor $M_0$ coincides with \eqref{M0}. As for the observables before, this form of the strong-coupling expansion is a reorganized version of the trans-series~\eqref{m-strong-Lambda}.

In~\cite{Basso:2009gh,Dorigoni:2015dha} it was shown that the perturbative part of the mass gap is related to the first non-perturbative correction of the cusp anomalous dimension. For a generic mixing angle, this relation reads
\begin{equation}\label{gamma-m-pert}
    \Gamma^{(\{0\},\{\})}=\frac{\left(x_0^+\right)^{2 a-1} }{8\pi g}{\Gamma(1-a)\over \Gamma(1+a)}\left[m^{(\{\},\{\})}\right]^2\,.
\end{equation}
Inverting this expression and using the explicit form for $\Gamma^{(\{0\},\{\})}$ from~\eqref{gamma-lead}, this relation implies for the mass gap
\begin{equation}\label{m-pert-xpm}
    m^{(\{\},\{\})}={\sqrt{8\pi g}\over \sqrt{S_1^{(\{0\},\{\})}-S_0^{(\{0\},\{\})}}}{\sqrt{Z_1^{(\{0\},\{\})}Z_0^{(\{\},\{\})}-Z_0^{(\{0\},\{\})}Z_1^{(\{\},\{\})}}\over  Z_0^{(\{\},\{\})}}\,.
\end{equation}
Following our conventions, the prefactor containing the Stokes constant ensures that the expression is automatically normalized as~\eqref{M-norm}. For the explicit values of the Stokes constants \cf~\eqref{Stokes-eg}.

Non-perturbative corrections for the mass gap can successively be obtained by acting with the Alien derivatives on the perturbative part $m^{(\{\},\{\})}$. Since we know the derivative's action on the determinants from~\eqref{AD-pm}, we can work out the Alien algebra for the mass gap as well. For the same reasons as for the tilted cusp anomalous dimension, we conjecture that the corresponding Alien derivatives are given by
\begin{equation}
\begin{aligned}\label{AD-m}
    &\Delta^+_jm^{(\delta^+,\delta^-)}=-{2i\sin\left(\alpha\right)}(N^+_j+1)\,(8\pi g)^{2a}m^{(\delta^+\cup\{j\},\delta^-)}\,,\\
    &\Delta^-_jm^{(\delta^+,\delta^-)}= {2i\sin\left(\alpha\right)}(N_j^-+1)\,(8\pi g)^{-2a}m^{(\delta^+,\delta^-\cup\{j\})} \,.
\end{aligned}
\end{equation}

Using the relation above, one can act on the perturbative part~\eqref{m-pert-xpm} with different Alien derivatives $\Delta_j^\pm$ to obtain different subleading contributions. As a further check of these relations, we used the Alien derivatives to calculate the $m^{(\delta^+,\delta^-)}$ appearing in the mass gap equivalent of~\eqref{Z-rel} and compared these to the $m^{(n,m)}$ obtained in~\eqref{eq:first-non-pert-massgap}, finding perfect agreement. For simplicity, we set the mixing angle to $a=1/4$ and to $a=1/(2\sqrt{2})$. The first value corresponds to the physical $\mathrm{O}(6)$ mass gap, while the irrationality of the second value ensures that the trans-series parameters do not mix, \ie~$\Lambda_+^{2n} \neq \Lambda_-^{2m}$ for all $n,m> 0$. The agreement between the two notations again suggests that our conjecture~\eqref{AD-m} and the underlying assumptions are valid and can be used to generate arbitrary non-perturbative corrections not only for the tilted cusp anomalous dimension but also for the tilted mass gap.

Moreover, also in the case of the mass gap we can prove the ratios~\eqref{m-ratio-Lambdapm}.
Similarly to~\eqref{first-AD} and~\eqref{ind-init}, acting on the perturbative part multiple times with the Alien derivative $\Delta^+_0$ and taking the ratio yields
\begin{equation}\label{M-ratios}
     {m^{(\{0^{(n+1)}\},\{\})}\over m^{(\{0^{(n)}\},\{\})}}=-{Z_0^{(\{0\},\{\})}\over Z_0^{(\{\},\{\})}}\,.
\end{equation}
Again, we understand $\{0^{(m)}\}$ as the set containing $m$-times the zeros. Again, we refer the reader to Appendix~\ref{App:Proof} for the proof.
Let us emphasize that this result, together with~\eqref{gamma-ratio} is in perfect agreement with equation (5.35) of~\cite{Dorigoni:2015dha}.

Similarly to the cusp anomalous dimension, we were unable to find any simple relation between different non-perturbative sectors other than~\eqref{M-ratios}.

Finally, we would like to point out an important consequence of the relation~\eqref{gamma-m-pert} between the perturbative corrections of the tilted cusp anomalous dimension and mass gap, as well as the universal structure~\eqref{AD-gamma} and~\eqref{AD-m} of the Alien derivatives. Starting from~\eqref{m-trans-xpm}, we parametrize the square of the mass gap as
\begin{equation}\label{m-sq-trans}
    m^2(\alpha)=M^2_0\sum_{ \delta^+, \delta^-}(8\pi g)^{2a \Delta}e^{i\alpha \Delta }e^{-8\pi g \left(\sum_{l\in  \delta^+}  x_l^++\sum_{j\in  \delta^- } x_j^-\right)}\left[m^2\right]^{( \delta^+, \delta^-)}\,.
\end{equation}

Recall equation~\eqref{gamma-m-pert} connecting the perturbative tilted mass gap with the first non-perturbative correction of the cusp~\cite{Basso:2009gh}.
Let us now act on this relation with Alien derivatives. Using the Alien algebra relations for the tilted cusp~\eqref{AD-gamma} and for the mass gap~\eqref{AD-m}, we arrive at
\begin{equation}
    \Gamma^{(\delta^+\cup\{0\},\delta^-)}=\frac{\left(x_0^+\right)^{2 a-1} }{8\pi g(N_0^++1)}{\Gamma(1-a)\over\Gamma(1+a)}\left[m^2\right]^{(\delta^+,\delta^-)}\,.
\end{equation}
Note that most of the prefactors introduced by the derivatives cancel and only the factor $(N_0^++1)$ appears in the denominator. Further, we can take out the additional Alien derivative $\Delta^+_0$ for the cusp, leading to
\begin{equation} \label{eq:general-connection-np-cusp-m}
   \Delta^+_0\Gamma^{(\delta^+,\delta^-)}=-2\pi i\frac{\left(8\pi g x_0^+\right)^{2 a-1}}{\Gamma (a) \Gamma (a+1)}\left[m^2\right]^{(\delta^+,\delta^-)}\,.
\end{equation}
We have tested these relations for the non-perturbative corrections to the mass gap to the same orders as presented in~\eqref{Z-rel}.

Moreover, we can now resum~\eqref{eq:general-connection-np-cusp-m}, resulting in the exact relation between the mass gap and the tilted cusp given as
\begin{equation} \label{eq:Deep-Conn}
    m^2(\alpha)={4ix_0^+\over \sin(\alpha)^3\cos(\alpha)}e^{-8\pi x_0^+ g}\Delta^+_0\Gamma_{\mathrm{cusp}}(\alpha)\,.
\end{equation}

\subsection{Physical results ($a=1/4$)} \label{Sec:O6-mass-gap}

Finally, for the reader's convenience, here we turn to the physical case with $a=1/4$ and summarize the results above in a concise way. In this case, the functions $\Gamma_{\mathrm{cusp}}(\alpha)$ and $m(\alpha)$ coincide with the cusp anomalous dimension of $\mathcal N=4$ SYM and the dynamically generated mass gap of the $\mathrm{O}(6)$ model, respectively.

For the physical value of the mixing angle, a detailed discussion was already given for the strong-coupling expansion of the Fredholm determinants $Z_0$ and $Z_1$ and the cusp anomalous dimension in~\cite{Boldis:2026rkb,Bajnok:2026xri}. Therefore, we only recall their most important properties and additional relations that were not yet presented in the previous works.

\paragraph{Fredholm determinants.} At strong coupling, the Fredholm determinant with $a=1/4$ and arbitrary integer $\ell$ is given as
\begin{equation}\label{trans-D-spec}
    Z_\ell=A_\ell \sum_{\delta^+,\delta^-}(8\pi g)^{\Delta\over 2}e^{{i\pi\over 4} \Delta }e^{-8\pi g \left(\sum_{l\in \delta^+}  x_l^++\sum_{j\in \delta^- } x_j^-\right)}Z_\ell^{(\delta^+,\delta^-)}\,,
\end{equation}
with $x_j^+=j+1/4$ and $x_j^-=j+3/4$ with $j\in\mathbb{N}_0$. As before, $\delta^+$ and $\delta^-$ are unordered sets of integer numbers, which possibly contain the same entry multiple times. Further, we have $\Delta=|\delta^+|-|\delta^-|$, where $|\delta^\pm|$ denotes the cardinality of the sets $\delta^\pm$. The explicit form of the constants $A_\ell$ for $a=1/4$ and $\ell=0,1$ can be found in equations (3.53) and (3.55) of~\cite{Boldis:2026rkb}.

As noted around~\eqref{scale-mixture}, the first two non-perturbative corrections arise solely from $\Lambda_-$ contributions. Starting from the third non-perturbative correction, the different sectors start to mix. 
This is of course due to the special value $a=1/4$, that aligns the different non-perturbative corrections. For example, each of the corrections with $(\delta^+,\delta^-)=(\{0\},\{2\})$, $(\{1\},\{1\})$, and $(\{2\},\{0\})$ contributes with the same exponential weight, namely at order $e^{-24\pi g}$.

The perturbative part $Z_\ell^{(\{\},\{\})}$ is given as a function of $a$ and the moments $\mathcal I_n$, \cf~\eqref{eq:Def-D_ell}, evaluated at $a=1/4$ and $\mathcal{I}_n=J_n$
\begin{equation}
    Z_\ell^{(\{\},\{\})}\equiv \mathcal{D}_\ell\left[a={1\over 4},\,\mathcal{I}_n=J_n\right]\,,
\end{equation}
where $J_n$ denotes the value of \eqref{In} for the physical mixing angle and is given by
\begin{equation}\label{I-spec}
    J_{2k+1}=2^{4k}\beta(2k)\,,
    \qquad J_{2k}=-(2^{4k-2}-2^{2k-1}-2)\zeta(2k-1)\,,
\end{equation}
for $k\geq1$. The value of $J_2$ is understood as the $k\to 1$ limit of the second formula, specifically $J_2=-6\log2$. The explicit perturbative form of $\mathcal{D}_\ell$ is given in~\eqref{eq:pertDet} and~\eqref{eq:pertDetCoeff}.

The non-perturbative corrections can be parametrized as in~\eqref{Z-np}. The Stokes constants $S_\ell^{(\delta^+,\delta^-)}$ are generated by~\eqref{Sp-fin} and~\eqref{Sm-fin} with $a=1/4$ and the initial condition $S^{(\{\},\{\})}=1$. The function $\mathcal{D}_\ell^{(\delta^+,\delta^-)}$ is a series in $1/g$ and given by the perturbative function $\mathcal{D}_\ell[a,\mathcal I_n]$ evaluated at $a=1/4-\Delta$ and with the shifted moments $\mathcal{I}_n=J_n^{(\delta^+,\delta^-)}$. The latter are given by
\begin{equation}\label{Is-a=1/4}
    J_n^{(\delta^+,\delta^-)}=J_n-(-1)^{n-1}\sum_{l\in\delta^+}\frac{2}{\left(l+\frac{1}{4}\right)^{n-1}}+\sum_{j\in\delta^-}\frac{2}{(j+\frac{3}{4})^{n-1}}\,.
\end{equation}
As discussed in Section~\ref{Sec:Universal-Resurgence-Determinant}, the Stokes constants $S_\ell^{(\delta^+,\delta^-)}$ vanish if either $\delta^+$ or $\delta^-$ contain the same entry multiple times. Therefore, the trans-series~\eqref{trans-D-spec} is at most first order in every exponential weight $e^{-8\pi g x_j^\pm}$.

Finally, since each series $\mathcal{D}_\ell^{(\delta^+,\delta^-)}$ is asymptotic, the non-perturbative corrections are also related by their Alien derivatives
\begin{equation}
\begin{aligned}\label{AD-Z-spec}
    &\Delta^+_jZ^{(\delta^+,\delta^-)}=-i\sqrt{2}(N^+_j+1)\,(8\pi g)^{1\over 2}Z^{(\delta^+\cup\{j\},\delta^-)}\,,\\
    &\Delta^-_jZ^{(\delta^+,\delta^-)}= i\sqrt{2}(N_j^-+1)\,(8\pi g)^{-{1\over 2}}Z^{(\delta^+,\delta^-\cup\{j\})}\,,
\end{aligned}
\end{equation}
where $N_j^+$ and $N_j^-$ denote the number of occurrences of the element $j$ in $\delta^+$ or $\delta^-$, respectively, which can be either zero or one in the case of the determinant. Furthermore, the Alien derivatives satisfy the commutativity relations~\eqref{AD-pm-algebra} and due to the vanishing of second order Stokes constants, the second order Alien derivatives vanish as well.

\paragraph{Cusp anomalous dimension.} The cusp anomalous dimension, defined in~\eqref{tilted-cusp-det}, is given as the ratio of two Fredholm determinants~\eqref{trans-D-spec} with $\ell=1$ and $\ell=0$. Therefore, the strong-coupling expansion of the cusp anomalous dimension is given by the trans-series
\begin{equation}\label{trans-cusp-spec}
    \Gamma_{\mathrm{cusp}}= 2g\sum_{ \delta^+, \delta^-}(8\pi g)^{ \Delta\over2}e^{{i\pi \over 4}\Delta }e^{-8\pi g \left(\sum_{l\in  \delta^+}  x_l^++\sum_{j\in  \delta^- } x_j^-\right)}\Gamma^{( \delta^+, \delta^-)}\,.
\end{equation}
The perturbative part of the trans-series is given by the ratio of the perturbative parts of the two determinants
\begin{equation}\label{cusp-pert}
    \Gamma^{(\{\},\{\})}={Z_1^{(\{\},\{\})}\over Z_0^{(\{\},\{\})}}\,,
\end{equation}
and its leading terms in the $1/g$-expansion can be obtained from~\eqref{tilt-cusp-pert} by setting $a=1/4$ and $\mathcal{I}_n=J_n$.

The different non-perturbative sectors of the cusp anomalous dimension are related by their Alien derivatives, \cf~Section~\ref{Sec:Tilted-mass-gap}. Accordingly, $\Gamma^{(\delta^+,\delta^-)}$ satisfy the same algebra as in~\eqref{AD-Z-spec}. As we have discussed above, the Alien derivative acts as a \emph{derivative}, \ie~we can apply the chain rule and the Leibniz rule to act on the composition, product, or ratio of determinants $Z_\ell^{(\delta^+,\delta^-)}$. Therefore, different non-perturbative corrections can be generated from the perturbative part~\eqref{cusp-pert}, by acting with the Alien derivatives $\Delta^\pm_j$, using~\eqref{AD-Z-spec} and finally substituting~\eqref{Z-np}. Notice that unlike in the case of the determinant, for the cusp anomalous dimension, higher powers of a given weight $e^{-8\pi g x_j^\pm}$ also contribute in the trans-series.

Moreover, it can be shown that the pure $\delta^+$ or pure $\delta^-$ sectors (so that the other set $\delta^-$ or $\delta^+$ is empty) satisfy the relations in~\eqref{gamma-ratio}. We present the explicit proof in Appendix~\ref{App:Proof}. 

For $a=1/4$, and using the specific values of the moments~\eqref{I-spec}, the ratios become
\begin{equation}
\begin{aligned}\label{cusp-ratio-spec}
    &{\Gamma^{(\{0^{(n+1)}\},\{\})}\over \Gamma^{(\{0^{(n)}\},\{\})}}=\frac{2 \Gamma \left(\frac{3}{4}\right)}{(8\pi g)\Gamma \left(\frac{1}{4}\right)}\left[1-\frac{9-6\log2}{2( 8\pi  g)}-\frac{56 K-333+108 (3-\log 2) \log 2}{8 (8\pi g)^2}+O\left(g^{-3}\right)\right] \,,\\
    &{\Gamma^{(\{\},\{0^{(n+1)}\})}\over \Gamma^{(\{\},\{0^{(n)}\})}}=-\frac{\sqrt{3} \Gamma \left(-\frac{3}{4}\right)}{(8 \pi  g)6 \Gamma \left(-\frac{1}{4}\right)}\bigg[1-\frac{25-54 \log 2}{6(8 \pi  g)}+\\
    &\phantom{xxxxxxxxxxxxxxxxxxxxxxxxxx} +\frac{2232 K+1525-180  (25-27 \log 2)\log 2}{72( 8\pi g)^2} +O\left(g^{-3}\right)\bigg] \,,
\end{aligned}
\end{equation}
with $n\geq 1$. The first line was derived by expanding the ratio of~\eqref{Z10} with $a=1/4$ and~\eqref{eq:pertDet} around $g=\infty$. Similar relations can be derived for sectors with $(\delta^+,\delta^-)=(\{i^{(n)}\},\{\})$ and $(\delta^+,\delta^-)=(\{\},\{i^{(n)}\})$.

\paragraph{$\mathrm{O}(6)$ mass gap.} Finally, the mass gap of the $\mathrm{O}(6)$ model at strong coupling is given by the trans-series~\eqref{m-trans-xpm} as
\begin{equation}
    m_{\mathrm{O}(6)}=\frac{\sqrt{2} (2\pi g)^{1\over 4} e^{-\pi  g}}{\Gamma \left(\frac{5}{4}\right)}\sum_{ \delta^+, \delta^-}(8\pi g)^{ \Delta\over2}e^{{i\pi \over 4}\Delta }e^{-8\pi g \left(\sum_{l\in \delta^+}  x_l^++\sum_{j\in \delta^- } x_j^-\right)}m^{( \delta^+, \delta^-)}\,.
\end{equation}
The perturbative part of the trans-series~\eqref{m-pert-xpm} becomes
\begin{equation}\label{m-pert-det-spec}
    m^{(\{\},\{\})}=\sqrt{{\pi g\Gamma\left({1\over 4}\right)\over \Gamma\left({3\over 4}\right)}}{\sqrt{Z_1^{(\{0\},\{\})}Z_0^{(\{\},\{\})}-Z_0^{(\{0\},\{\})}Z_1^{(\{\},\{\})}}\over  Z_0^{(\{\},\{\})}}\,.
\end{equation}
The leading behavior of the perturbative part is explicitly given in~\eqref{m-pert-spec}.

The non-perturbative coefficients are related by the Alien derivatives which act for the different sectors according to~\eqref{AD-Z-spec}. Using~\eqref{AD-Z-spec} and~\eqref{Z-np} all non-perturbative corrections can be easily generated from the perturbative sector.

Similarly to the cusp anomalous dimension, the ratios of non-perturbative corrections $m^{(\{(0)^{(n)}\},\{\})}$
satisfy the general rule~\eqref{M-ratios}. Explicitly, the expansion in $1/g$ of these ratios is the same as in the first line of~\eqref{cusp-ratio-spec}, that is
\begin{equation}
    {m^{(\{0^{(n+1)}\},\{\})}\over m^{(\{0^{(n)}\},\{\})}}=\frac{2 \Gamma \left(\frac{3}{4}\right)}{(8\pi g)\Gamma \left(\frac{1}{4}\right)}\left[1-\frac{9-6\log2}{2( 8\pi  g)}-\frac{56 K-333+108 (3-\log 2) \log 2}{8 (8\pi g)^2}+O\left(g^{-3}\right)\right]\,.
\end{equation}
Note, however, that in this case we have $n\geq 0$. This relation, together with~\eqref{gamma-ratio},~\eqref{M-ratios} and~\eqref{cusp-ratio-spec} generalize the formulae and conjectures (5.33)-(5.35) of~\cite{Dorigoni:2015dha}.

Finally, let us turn to the deeper connection~\eqref{eq:Deep-Conn} of the strong-coupling expansions for the cusp anomalous dimension and the mass gap of the $\mathrm{O}(6)$ model. At strong coupling, we write the square of the mass gap as
\begin{equation}
    m^2_{\mathrm{O}(6)}=\frac{2 \sqrt{2\pi g}\, e^{-2\pi  g}}{\Gamma^2 \left(\frac{5}{4}\right)}\sum_{ \delta^+, \delta^-}(8\pi g)^{ \Delta\over2}e^{{i\pi \over 4}\Delta }e^{-8\pi g \left(\sum_{l\in \delta^+}  x_l^++\sum_{j\in \delta^- } x_j^-\right)}\left[m^2\right]^{( \delta^+, \delta^-)}\,.
\end{equation}
For each non-perturbative order $(\delta^+,\delta^-)$ the non-perturbative corrections of the cusp anomalous dimension and the squared mass gap satisfy
\begin{equation}
\begin{aligned}
    \Delta^+_0\Gamma^{(\delta^+,\delta^-)} &=-i\sqrt{2}(N_0^++1)(8\pi g)^{1\over2}\Gamma^{(\delta^+\cup\{0\},\delta^-)} \\
    &=-i\sqrt{2}(8\pi g)^{-{1\over2}}\frac{8 \Gamma \left({3\over 4}\right)}{\Gamma \left({1\over 4}\right)}\left[m^2\right]^{(\delta^+,\delta^-)}\,.
    \end{aligned}
\end{equation}
This relation holds to all orders and is the generalization of (4.21) of~\cite{Basso:2009gh} and (3.31)  of~\cite{Dorigoni:2015dha}.
Resumming both sides, we arrive at the formal relation between the two quantities
\begin{equation}
    m^2_{\mathrm{O}(6)}=4 ie^{-2\pi g}\Delta^+_0\Gamma_{\mathrm{cusp}}\,.
\end{equation}
The equations above are formal, all-order strong-coupling relations and they connect two different AdS/CFT observables, namely the $\mathrm{O}(6)$ mass gap and the cusp anomalous dimension of $\mathcal{N}=4$ SYM theory, emerging from two different sides of the holographic duality. The last relation is understood as the Alien derivative acts on the strong-coupling non-perturbative sectors $\Gamma^{(\delta^+,\delta^-)}$. Using the formalism and ingredients that we presented in this article, it can be explicitly checked to hold order-by-order.

Based on these results, a beautifully compact notation for the cusp anomalous dimension and the $\mathrm{O}(6)$ mass gap is given in~\cite{Bajnok:2026xri}.

\section{Discussion and Conclusions} \label{Sec:Discussion}
The main goal of this paper was to explore the universal strong-coupling structure of the dynamically generated mass gap in the $\mathrm{O}(6)$ model. It is known~\cite{Basso:2009gh} that the perturbative part of the mass gap is connected to the first non-perturbative correction of the cusp anomalous dimension of $\mathcal{N}=4$ SYM. Using recent results for the latter~\cite{Bajnok:2025mxi,Boldis:2026rkb,Bajnok:2026xri}, we can express the mass gap at large values of the~'t~Hooft coupling in terms of non-perturbative corrections of Fredholm determinants with matrix Bessel kernel, which possess a well-understood strong-coupling structure. 

With this in mind, we reviewed the connection between the tilted cusp and the BES equation as well as Fredholm determinants in Section~\ref{Sec:tiltedCusp}. In Section~\ref{Sec:MethodDiffEq} we presented the established method of differential equations. Since the cusp anomalous dimension is related to the behavior of the Bethe root density at the origin, we also considered the solution to the set of differential equations at this special point.
Afterwards, we introduced a one-parameter deformation of the $\mathrm{O}(6)$ mass gap in Section~\ref{Sec:TiltedMassGap}, referred to as the tilted mass gap, which becomes the physical observable $m_{\mathrm{O}(6)}$ for $a=1/4$. In the same Section, we also discussed its connection to the tilted cusp.

Finally, we turned to the universal structure of the strong-coupling expansions in Section~\ref{Sec:Universal-Structure}. Firstly, we reviewed the findings~\cite{Bajnok:2025mxi,Boldis:2026rkb} for Fredholm determinants. Most importantly, we discussed the \emph{shift relations}~\eqref{D-rule} for the large-$g$ expansion at a given non-perturbative order as well as the \emph{recurrence relations}~\eqref{Sp-fin},~\eqref{Sm-fin} for the Stokes constants. The \emph{Alien derivatives}~\eqref{AD-pm} allowed us to generate all the non-perturbative sectors from the perturbative part.
Secondly, the tilted cusp can be written as a ratio of two determinants. We reviewed the resulting Alien derivatives~\cite{Boldis:2026rkb} and proved the relation for the ratio of two successive non-perturbative corrections to the tilted cusp that was first observed in~\cite{Dorigoni:2015dha}.
Equipped with the formalism, we turned to the tilted mass gap in Section~\ref{Sec:Tilted-mass-gap}. Knowing how the perturbative part of the mass gap can be expressed through the first non-perturbative correction of the cusp, we devised the corresponding Alien derivatives~\eqref{AD-m}. A relation also exists for the mass gap's ratio of two successive non-perturbative corrections~\eqref{M-ratios}, which we proved in Appendix~\ref{App:Proof}. Finally, we can derive the exact relation~\eqref{eq:Deep-Conn} between the mass gap and the cusp anomalous dimension.
In Section~\ref{Sec:O6-mass-gap} we summarized our results for the physical case of the mass gap in the $\mathrm{O}(6)$ model (for~$a=1/4$).

In conclusion, in this article, we have presented how the strong-coupling behavior of the mass gap is determined from the strong-coupling expansion of Fredholm determinants. The resurgence structure of the latter is well understood~\cite{Bajnok:2025lji,Boldis:2026rkb}.
Using the Alien calculus the full non-perturbative mass gap can be generated. Our discussion gives a nontrivial all-order relation between two observables from different regimes of the AdS/CFT correspondence.

It would be very interesting to establish a connection with other observables, such as the energy density of the $\mathrm{O}(6)$ model. An expression for the perturbative part in terms of Fredholm determinants would already be sufficient, as we could then use the Alien algebra and shift relations to compute physical observables at strong coupling up to arbitrary exponential orders. The connection between the cusp and the mass gap~\eqref{gamma-m-pert} was our starting point in this work. Relations like this seem to hint at a deeper structure within the AdS/CFT correspondence. Finally, it would be fascinating to understand, whether the $\mathrm{O}(6)$ non-linear sigma model can be deformed in such a way, that \eqref{m-gen-fin} arises in a natural way.\footnote{We thank Daniele Dorigoni for discussions related to this point.}

Let us also add that the non-perturbative structure of $\mathrm{O}(N)$ models was investigated in~\cite{Bajnok:2025mxi}. There, a set of exact differential equations was derived from the TBA equations, connecting observables and the Bethe root densities. Starting from the perturbative part of \emph{one} known observable, the perturbative contribution of \emph{all} observables can be constructed. Moreover, this ODE system provides a basis of building blocks that allow the construction of arbitrary non-perturbative corrections as a sum over possible contributions to a given power of the trans-series parameter. It would be fascinating to understand how this method in the case of the $\mathrm{O}(6)$ model is connected to the universal structure of the Fredholm determinants that we used here.

\section*{Acknowledgments}

We thank Zoltan Bajnok, Gregory Korchemsky, Alessandro Testa and Daniele Dorigoni for fruitful discussions and comments related to this work.
The research was supported by the Doctoral Excellence Fellowship Programme
funded by the National Research Development and Innovation Fund of
the Ministry of Culture and Innovation and the Budapest University
of Technology and Economics, under a grant agreement with the National
Research, Development and Innovation Office (NKFIH). It was also supported by the grant NKKP Advanced 152467. 
DlP is grateful for the funding through a Feodor Lynen Fellowship by the Alexander von Humboldt Foundation.

\appendix

\section{The resolvent and $q_\pm$} \label{App:A}

In this appendix, we prove some of the relations of the main text. This Appendix is based on the concepts and notations of~\cite{Bajnok:2024bqr}. For simplicity, we choose $\ell=0$.

Following~\cite{Bajnok:2024bqr}, we start by defining the two-component states
\begin{equation} \label{Psi}
\ket{\Psi_{2n-1}} =\lr{\ket{\psi_{2n-1}}\atop 0}\,,\qqqquad \ket{\Psi_{2n}} =\lr{0 \atop \ket{\psi_{2n}}}\,,
\end{equation}
with $\psi_{n}(x)=\langle x|\psi_n\rangle$ given by Bessel functions
\begin{equation}\label{psi-def}
    \psi_n(x)=\sqrt{n}{J_n(\sqrt{x})\over\sqrt{x}}\,.
\end{equation}
These states satisfy the orthogonality condition $\langle\Psi_n|\Psi_m\rangle=\delta_{nm}$.
It is convenient to reshuffle the columns and rows of the infinite matrix $\mathbb{K}(\alpha)$, such that its elements are given by
\begin{equation}
    \mathbb K_{nm}(\alpha)=\left\langle \Psi_n|\mathcal X|\Psi_m\right\rangle\,.
\end{equation}
The operator $\bm{\mathcal{X}}$ acts on a test function $f$ as
\begin{equation}
    \bm{\mathcal X} f(x) = \chi\Big(\frac{\sqrt{x}}{2g}\Big) U f(x)\,,\qquad U=\cos\alpha \left(\begin{array}{rr}\cos\alpha & \sin\alpha \\ -\sin\alpha & \cos\alpha\end{array}\right)\,.
\end{equation}

The advantage of this is that now we can rewrite $Z_0(\alpha)$ as a Fredholm determinant
\begin{equation}
    Z_0(\alpha)=\det\left(1+\bm{\mathcal{HX}}\right)\,,
\end{equation}
where the operator $\bm{\mathcal{H}}$ is given as
\begin{equation}
\bm{\mathcal{H}}=\sum_{n\geq1}|\Psi_n\rangle\langle\Psi_n|\,,
\end{equation}
The kernel of this operator is a diagonal two-by-two matrix
\begin{equation}
    \mathcal{H}(x,y)=\begin{pmatrix}
        \sum_{n\geq1}\psi_{2n-1}(x)\psi_{2n-1}(y)&0\\
        0&\sum_{n\geq1}\psi_{2n}(x)\psi_{2n}(y)
    \end{pmatrix}\,.
\end{equation}
Notice that due to the orthogonality of the states $|\Psi_n\rangle$, the operator $\bm{\mathcal{H}}$ satisfies
\begin{equation}\label{project}
    \bm{\mathcal H}|\Psi_n\rangle=|\Psi_n\rangle\,,\qquad \bm{\mathcal H}\bm{\mathcal H}=\bm{\mathcal H}\,.
\end{equation}

\subsection*{Resolvent}

Consider the following matrix of functions
\begin{equation}
    \gamma(x,y)=\begin{pmatrix}
        \gamma_{11}(x,y)&\gamma_{12}(x,y)\\
        \gamma_{21}(x,y)&\gamma_{22}(x,y)
    \end{pmatrix}=\sum_{n,m\geq 1}\left\langle x|\Psi_n\right\rangle\left[{1\over1+\mathbb K(\alpha)}\right]_{nm}\left\langle\Psi_m|y\right\rangle\,.
\end{equation}
For later purposes, it is convenient to explicitly write its elements, which read
\begin{equation}
\begin{aligned}
\label{eq:gamma_matrix}
\gamma_{11}(x,y)= & \sum_{n,m\geq1}\psi_{2n-1}(x)\psi_{2m-1}(y)\left(\frac{1}{1+\mathbb{K}(\alpha)}\right)_{2n-1,2m-1}\,,\\
\gamma_{12}(x,y)= & \sum_{n,m\geq1}\psi_{2n-1}(x)\psi_{2m}(y)\left(\frac{1}{1+\mathbb{K}(\alpha)}\right)_{2n-1,2m}\,, \\
\gamma_{21}(x,y)= & \sum_{n,m\geq1}\psi_{2n}(x)\psi_{2m-1}(y)\left(\frac{1}{1+\mathbb{K}(\alpha)}\right)_{2n,2m-1}\,, \\
\gamma_{22}(x,y)= & \sum_{n,m\geq1}\psi_{2n}(x)\psi_{2m}(y)\left(\frac{1}{1+\mathbb{K}(\alpha)}\right)_{2n,2m}\,.
\end{aligned}
\end{equation}

It is straightforward to show that $\gamma(x,y)$ is the kernel of the resolvent operator with
\begin{equation}
    \bm{\gamma}={1\over 1+\bm{\mathcal{HX}}}\bm{\mathcal H}
\end{equation}
From this representation, it immediately follows that
\begin{equation}
    \bm{\gamma}(1+\bm{\mathcal{XH}})=\bm{\mathcal H}\,.
\end{equation}
Multiplying both sides of the equation by $|\Psi_n\rangle$ we get
\begin{equation}
    \bm{\gamma}(1+\bm{\mathcal{XH}})|\Psi_n\rangle=\bm{\mathcal H}|\Psi_n\rangle\,.
\end{equation}
Using~\eqref{project}, this leads to
\begin{equation}
    \bm{\gamma}(1+\bm{\mathcal{X}})|\Psi_n\rangle=|\Psi_n\rangle\,,
\end{equation}
which is equivalent to the integral equation
\begin{equation}
    \int dy\gamma(x,y)\left[1+U\chi\left({\sqrt{y}\over 2g}\right)\right]\Psi_n(y)=\Psi_n(x)\,.
\end{equation}
This yields a pair of infinite sets of integral equations for the components of $\gamma(x,y)$. For odd and even $n$, the first component of the above equation looks as
\begin{equation}
\begin{aligned}\label{inteq-gamma-mat}
    &\int_0^\infty dy\, \psi_{2n-1}\left(y\right)\left[\left(1+\cos^2(\alpha)\chi\left({\sqrt{y}\over2g}\right)\right)\gamma_{11}(x,y)-\sin(\alpha)\cos(\alpha)\chi\left({\sqrt{y}\over2g}\right)\gamma_{12}(x,y)\right]=\psi_{2n-1}(x)\,,\\
    &\int_0^\infty dy\, \psi_{2n}\left(y\right)\left[\sin(\alpha)\cos(\alpha)\chi\left({\sqrt{y}\over2g}\right)\gamma_{11}(x,y)+\left(1+\cos^2(\alpha)\chi\left({\sqrt{y}\over2g}\right)\right)\gamma_{12}(x,y)\right]=\psi_{2n}(x)\,.
\end{aligned}
\end{equation}

The functions in~\eqref{eq:gamma_matrix} are related to $\gamma_\pm(x)$ by
\begin{equation}
\begin{aligned}\label{gamma-pm-def}
    \gamma_-(x)&=4gx\lim_{y\to 0}\gamma_{11}\left((2gy)^2,(2gx)^2\right)\,,\\
    \gamma_+(x)&=4gx\lim_{y\to 0}\gamma_{12}\left((2gy)^2,(2gx)^2\right)\,.
\end{aligned}
\end{equation}
With a change of variables in~\eqref{eq:gamma_matrix} and using the definition~\eqref{psi-def} for the $\psi_n(x)$ states, we take the $x\to0$ limits of the first two lines and find the following expressions as a product of Bessel functions and the resolvent
\begin{equation}
\begin{aligned}
    &\gamma_-(x)=\sum_{m\geq1}\sqrt{2m-1}J_{2m-1}(2gx)\left(\frac{1}{1+\mathbb{K}(\alpha)}\right)_{1,2m-1}\\
    &\gamma_+(x)=\sum_{m\geq1}\sqrt{2m}J_{2m}(2gx)\left(\frac{1}{1+\mathbb{K}(\alpha)}\right)_{1,2m}\,.
\end{aligned}
\end{equation}
Repeating the same procedure for the equations of~\eqref{inteq-gamma-mat}, one finds that $\gamma_\pm(x)$ satisfy the integral equations in~\eqref{int-elem-zero}.

\subsection*{Relation to $q_\pm(x)$}

We start by defining the functions
\begin{equation}
    Q(x)=\bra{x}{1\over 1+\bm{\mathcal H} \bm{\mathcal X}} \ket{{\mathit\Phi}}\rangle\,,\qquad P(x)=\bra{x}{1\over 1+\bm{\mathcal H} \bm{\mathcal X}} \ket{\dot{\mathit\Phi}}\rangle\,,
\end{equation}
as well as the expectation value
\begin{equation}\label{u}
u= - \langle\bra{\mathit\Phi} \bm{\mathcal X} {1\over 1+ \bm{\mathcal H} \bm{\mathcal X}}\ket{\mathit\Phi}\rangle
\,,
\end{equation}
which is conventionally called the potential. The states are defined by
\begin{equation}\label{Phi-ell}
{\ket{\mathit\Phi}\rangle} =  \left(\begin{array}{cc}  \ket{\phi_0} & 0 \\0 & \ket{\phi_{1}} \end{array}\right)\,,\qqqquad
\phi_\ell(x) = J_{\ell}(\sqrt x)\,,
\end{equation}
and $\ket{\dot{\mathit\Phi}}\rangle\equiv x\partial_x\ket{\mathit\Phi}\rangle$. It is important to note that $Q(x)$ is a two-by-two matrix of real functions, satisfying the symmetry relation
 \begin{equation}\label{Q-sym}
Q(x|-\alpha)=\sigma_3Q(x|\alpha) \sigma_3\,,
\end{equation}
where $\sigma_3$ is the Pauli matrix. 

The potential $u$ is directly related to the determinant~\eqref{F-def} for $\ell=0$ via
\begin{equation}\label{eq1}
g\partial_g \log Z_0(\alpha)=-\frac12 \tr u\,.
\end{equation}
Further, it satisfies the symmetry properties
\begin{equation}\label{u-symm}
    u(\alpha)=\sigma_3u(\alpha)^t\sigma_3\,, \qquad u(\alpha)=u^t(-\alpha)\,.
\end{equation}
Let us now define the two-by-two matrix
\begin{equation}
    w=u-g\partial_gu\,.
\end{equation}
The functions $W_0$ and $W_\pm$ introduced in Section~\ref{Sec:MethodDiffEq} are given by
\begin{equation}
    W_0=\frac{1}{2}\left(w_{11}+w_{22}-1\right)\,, \qquad
    W_\pm=-\frac{1}{2}\left(w_{11}-w_{22}+1\right)\pm w_{12}\,.
\end{equation}
Finally, using~\eqref{eq1}, the function $W_0(g)$ is related to the observable by
\begin{equation}
    g^2 \partial_g^2 \, \log{Z_0(g|\alpha)} = W_0(g) +\frac{1}{2}\,. \label{eq:Z-and-W}
\end{equation}

As shown in Appendix A and B of~\cite{Bajnok:2024bqr}, the functions $Q(x)$ and $P(x)$ satisfy the differential equation
\begin{equation}\label{diffeq-PQ}
    \lr{x\partial_x+\frac12 g\partial_g} Q(x)=\frac14Q(x) u+ P(x)\,.
\end{equation}
Moreover, they are related to the resolvent via
\begin{equation}\label{gam-P-Q} 
\gamma(x,y)
 ={Q(x) \sigma_3 P^t(y) \sigma_3-P(x) \sigma_3 Q^t(y) \sigma_3\over x-y}\,,
\end{equation}
where the superscript `$t$' denotes the transpose of the two-by-two matrix. Using~\eqref{diffeq-PQ}, one can express $P(x)$ in terms of $Q(x)$ and substitute it into~\eqref{gam-P-Q} to express the resolvent in terms of $Q(x)$ and its derivatives as
\begin{equation}\label{gamma-Q}
    \gamma(x,y)={Q(x)\sigma_3 \left(2y\partial_y+g\partial_g\right)Q^t(y)\sigma_3-\left(2x\partial_x+g\partial_g\right)Q(x)\sigma_3 Q^t(y)\sigma_3\over 2(x-y)}\,.
\end{equation}
Since $\gamma(x,y)$ and $Q(x)$ are two-by-two matrices, this relation gives four equations between the elements of the resolvent $\gamma(x,y)$ and the function matrix $Q(x)$.

In the next step we define the functions $q_\pm(x)$ as a combination of elements of $Q(x)$ given by
\begin{equation}\label{q-Q}
q_\pm (x)=e^{-i\alpha/2} \left[ Q_{11}((2gx)^2) \pm Q_{12}((2gx)^2) + i Q_{21}((2gx)^2)\pm  i Q_{22}((2gx)^2)\right]\,.
\end{equation}
Now $q_\pm(x)$ are complex valued and by~\eqref{Q-sym} they satisfy the symmetry properties
\begin{equation}
\label{q-pma-conj}
    q_+(-x|-\alpha)=e^{i\alpha}q_-(x|\alpha)\,,\qquad \bar q_\pm(x|\alpha)=e^{i\alpha}q_\pm(-x|\alpha)
\end{equation}

Inverting~\eqref{q-Q}, substituting the matrix elements of $Q(x)$ into~\eqref{gamma-Q}, taking the limit $x\to 0$ and finally using the definitions~\eqref{gamma-pm-def} we find two relations between the functions $\gamma_{\pm}(x)$ and $q_\pm(x)$, which read
\begin{equation}
\begin{aligned}
    \label{gammapm-qpm}
    \gamma_+(x)=\frac{e^{i\alpha}}{8ix}&\left[q_+(0)\left(\partial_gq_-(x)-\partial_gq_-(-x)\right)+q_-(0)\left(\partial_gq_+(x)-\partial_gq_+(-x)\right)-\right.\\
    &-\partial_gq_+(0)\left(q_-(x)-q_-(-x)\right)-\partial_gq_-(0)\left(q_+(x)-q_+(-x)\right)\left.\right]\\
    \gamma_-(x)=-\frac{e^{i\alpha}}{8x}&\left[q_+(0)\left(\partial_gq_-(x)+\partial_gq_-(-x)\right)+q_-(0)\left(\partial_gq_+(x)+\partial_gq_+(-x)\right)-\right.\\
    &-\partial_gq_+(0)\left(q_-(x)+q_-(-x)\right)-\partial_gq_-(0)\left(q_+(x)+q_+(-x)\right)\left.\right]\,.
\end{aligned}
\end{equation}
Here we have used~\eqref{q-pma-conj}. Finally, by definition~\eqref{capital-G}, the expression~\eqref{gamma-q-rel2} follows immediately.

\section{Non-perturbative corrections at the origin}\label{Sec:AppB}

In this appendix, we provide some non-perturbative corrections for the functions $c_+$ and $c_-$ at strong coupling, which are related to $q_\pm(x)$ at $x=0$, see equation~\eqref{q0-ansatz}.

Some $1/g$ coefficients of the perturbative part of $c_+$ are presented in~\eqref{cp-sol}.

The first few non-perturbative corrections for the function $c_+(g)$ are
\begin{equation}
\begin{aligned}
    c_+^{(1,0)}&(g)=-4^{-2 a} (1-2 a)^{2 a} \Gamma (1-a)\times\\
    &\times\left[1+\frac{a (\mathcal{I}_2+2)}{8\pi g}-\frac{a \left(4 (1-a)+(1-a) (1-2 a)^2 (\mathcal{I}_2+2)^2-(1-3 a) (1-2a)^2 \mathcal{I}_3\right)}{2 (1-2 a)^2 (8\pi g)^2}+O\left(g^{-3}\right)\right]\,,\\
    c_+^{(2,0)}&(g)=-\frac{2^{1-6 a} (1-2 a)^{4 a-1} \Gamma (1-a)^2}{8\pi g\Gamma (a)}\left[1+\frac{a^2 (6 \mathcal{I}_2+8)-5 a (\mathcal{I}_2+2)+\mathcal{I}_2+4}{(2 a-1) 8\pi g}+O\left(g^{-2}\right)\right]\,,\\
    c_+^{(3,0)}&(g)=-\frac{4^{1-4 a} (1-2 a)^{6 a-2} \Gamma (1-a)^3}{(8\pi g)^2\Gamma (a)^2}\left[1+O\left(g^{-1}\right)\right]\,,\\
    c_+^{(0,1)}&(g)=\frac{2 (1+2 a)^{-2 a-1} \Gamma (1+a)^2}{(8\pi g)\Gamma (-a)}\left[1-\frac{(3 a+1) \mathcal{I}_2+2 (a+1)}{8\pi g}+O\left(g^{-2}\right)\right]\,,\\
    c_+^{(1,1)}&(g)=\frac{1}{2} (1-2 a)^{1+2 a} (2+4 a)^{-2 a} \Gamma (1+a)\times\\
    &\times\left[1-\frac{a \mathcal{I}_2}{8\pi g}+\frac{a \left(32 a^2+(1+a) \left(1-4 a^2\right)^2 \mathcal{I}_2^2+(3 a+1)\left(1-4 a^2\right)^2 \mathcal{I}_3\right)}{2 \left(1-4 a^2\right)^2 (8\pi g)^2}+O\left(g^{-3}\right)\right]\,.
\end{aligned}
\end{equation}

While the first few non-perturbative corrections for the function $c_-$ are
\begin{equation}
\begin{aligned}
    c_-^{(1,0)}&(g)=\frac{2 (1-2 a)^{2 a-1} \Gamma (1-a)^2}{(8\pi g)\Gamma (a)}\left[1-\frac{2(1-a)+(1-3 a) \mathcal{I}_2}{8\pi g}+O\left(g^{-2}\right)\right]\,,\\
    c_-^{(2,0)}&(g)=\frac{4^{1-a} (1-2 a)^{4 a-2} \Gamma (1-a)^3}{(8\pi g)^2\Gamma (a)^2}\left[1+O\left(g^{-1}\right)\right]\,,\\
    c_-^{(3,0)}&(g)=O\left(g^{-3}\right)\,,\\
    c_-^{(0,1)}&(g)=-4^{2 a} (1+2 a)^{-2 a} \Gamma (1+a)\times\\
    &\times\left[1-\frac{a (\mathcal{I}_2+2)}{8\pi g}+\frac{a \left(4 (1+a)+(1+a) (1+2 a)^2 (\mathcal{I}_2+2)^2+(1+3 a) (1+2a)^2 \mathcal{I}_3\right)}{2 (1+2 a)^2 (8\pi g)^2}+O\left(g^{-3}\right)\right]\,,\\
    c_-^{(1,1)}&(g)=\frac{1}{2} (1+2 a)^{1-2a} (2-4a)^{2 a} \Gamma (1-a)\times\\
    &\times\left[1+\frac{a \mathcal{I}_2}{8\pi g}-\frac{a \left(32 a^2+(1-a) \left(1-4 a^2\right)^2 \mathcal{I}_2^2+ (3 a-1)\left(1-4 a^2\right)^2 \mathcal{I}_3\right)}{2 \left(1-4 a^2\right)^2 (8\pi g)^2}+O\left(g^{-3}\right)\right]\,.
\end{aligned}
\end{equation}

Further coefficients can be obtained using the symmetry relation
\begin{equation}
    c^{(n,m)}_-(a)=c_+^{(m,n)}(-a)\,.
\end{equation}

\section{Proof for the ratios} \label{App:Proof}

In this section, we prove the relations~\eqref{gamma-ratio} and~\eqref{M-ratios}. In order to do so, we will use the following three assumptions:
\begin{enumerate}
    \item The perturbative part of the cusp anomalous dimension and the mass gap is given by~\eqref{gamma-pert} and~\eqref{m-pert-xpm} in terms of the Fredholm determinants.
    \item The Alien derivatives act as derivatives on simple functions and obey the Leibniz rule and the chain rule.
    \item The Alien algebra in~\eqref{AD-gamma} and~\eqref{AD-m} holds.
\end{enumerate}

For simplicity, we only present the proof for the non-perturbative sectors $\Gamma^{(\{0^{(j)}\},\{\})}$ with $j\geq 1$ of the tilted cusp anomalous dimension. For $\Gamma^{(\{i^{(j)}\},\{\})}$, $\Gamma^{(\{\},\{i^{(j)}\})}$ and for the corrections $m^{(\{0^{(j)}\},\{\})}$ with $j\geq 0$ of the tilted mass gap, the proof works analogously.

According to~\eqref{ind-init}, the following relation holds for $n=1,2$
\begin{equation}\label{ind-cond}
    \Gamma^{(\{0^{(n)}\},\{\})}=\left[-{Z_0^{(\{0\},\{\})}\over Z_0^{(\{\},\{\})}}\right]^{n-1}\Gamma^{(\{0\},\{\})}\,.
\end{equation}
Now we prove that if this relation holds for $n$, then it also holds for $n+1$, and hence by induction~\eqref{ind-cond} is valid for arbitrary $n\geq 1$.

We begin by acting on $\Gamma^{(\{0^{(n)}\},\{\})}$ with the Alien derivative $\Delta^+_0$. According to the second assumption above, this yields
\begin{equation}
    \label{proof1}
    \Delta^+_0\Gamma^{(\{0^{(n)}\},\{\})}=(-1)^{n-1}\Delta^+_0\left[{Z_0^{(\{0\},\{\})}\over Z_0^{(\{\},\{\})}}\right]^{n-1}\Gamma^{(\{0\},\{\})}+(-1)^{n-1}\left[{Z_0^{(\{0\},\{\})}\over Z_0^{(\{\},\{\})}}\right]^{n-1}\Delta^+_0\Gamma^{(\{0\},\{\})}\,.
\end{equation}

The derivative in the first term leads to
\begin{equation}
\begin{aligned}\label{T1}
    \Delta^+_0\left[{Z_0^{(\{0\},\{\})}\over Z_0^{(\{\},\{\})}}\right]^{n-1}=&(n-1)\left[{Z_0^{(\{0\},\{\})}\over Z_0^{(\{\},\{\})}}\right]^{n-2}\Delta^+_0{Z_0^{(\{0\},\{\})}\over Z_0^{(\{\},\{\})}}\\
    =&(n-1)\left[{Z_0^{(\{0\},\{\})}\over Z_0^{(\{\},\{\})}}\right]^{n-2}{\left[\Delta^+_0Z_0^{(\{0\},\{\})}\right]Z_0^{(\{\},\{\})}-\left[\Delta^+_0Z_0^{(\{\},\{\})}\right]Z_0^{(\{0\},\{\})}\over  \left[Z_0^{(\{\},\{\})}\right]^2}\\
    =&2i\sin(\pi a)(8\pi g)^{2a}(n-1)\left[{Z_0^{(\{0\},\{\})}\over Z_0^{(\{\},\{\})}}\right]^{n}
\end{aligned}
\end{equation}
In the first line, we used the power rule and in the second line the quotient rule for the derivative. In the last relation, we used~\eqref{AD-sqr} (\ie~$Z_0^{(\{0,0\},\{\})}=0$) and~\eqref{AD-pm} on $Z_0^{(\{\},\{\})}$.

Now we turn to the second term of~\eqref{proof1}. According to~\eqref{AD-gamma}, the Alien derivative of $\Gamma^{(\{0\},\{\})}$ reads
\begin{equation}
    \Delta^+_0\Gamma^{(\{0\},\{\})}=-{4i\sin\left(\pi a\right)}\,(8\pi g)^{2a}\Gamma^{(\{0,0\},\{\})}\,.
\end{equation}
Here we used the fact that $N_0^+=1$ for $\delta^+=\{0\}$. Following~\eqref{ind-init}, this becomes
\begin{equation}\label{T2}
    \Delta^+_0\Gamma^{(\{0\},\{\})}={4i\sin\left(\pi a\right)}\,(8\pi g)^{2a}{Z_0^{(\{0\},\{\})}\over Z_0^{(\{\},\{\})}}\Gamma^{(\{0\},\{\})}\,.
\end{equation}

Using~\eqref{T1} and~\eqref{T2}, the Alien derivative in~\eqref{proof1} evaluates to
\begin{equation}
    \Delta_0^+\Gamma^{(\{0^{(n)}\},\{\})}=-2i\sin(\pi a)(8\pi g)^{2a}(n+1)\left[-{Z_0^{(\{0\},\{\})}\over Z_0^{(\{\},\{\})}}\right]^{n}\Gamma^{(\{0\},\{\})}\,.
\end{equation}
According to~\eqref{AD-gamma}, this Alien derivative is proportional to $\Gamma^{(\{0^{(n+1)}\},\{\})}$. Since for $\delta^+=\{0^{(n)}\}$ we have $N_0^+=n$ this implies
\begin{equation}
    \Gamma^{(\{0^{(n+1)}\},\{\})}=\left[-{Z_0^{(\{0\},\{\})}\over Z_0^{(\{\},\{\})}}\right]^{n}\Gamma^{(\{0\},\{\})}\,.
\end{equation}
Hence,~\eqref{ind-cond} holds for arbitrary $n$. Finally, dividing $\Gamma^{(\{0^{(n+1)}\},\{\})}$ by $\Gamma^{(\{0^{(n)}\},\{\})}$, we get
\begin{equation}
    {\Gamma^{(\{0^{(n+1)}\},\{\})}\over \Gamma^{(\{0^{(n)}\},\{\})}}=-{Z_0^{(\{0\},\{\})}\over Z_0^{(\{\},\{\})}}\,,
\end{equation}
which completes our proof.

\bibliographystyle{JHEP}    
\bibliography{references}      
     
\end{document}